\documentclass[nofootinbib,preprintnumbers,prd,superscriptaddress,twocolumn,showpacs]{revtex4-1}

\usepackage{amsmath}
\usepackage{amsfonts}
\usepackage{amssymb}
\usepackage{graphicx}
\usepackage[titletoc]{appendix}
\usepackage{color}
\usepackage{hyperref}
\usepackage{cleveref}
\usepackage[rightcaption]{sidecap}
\usepackage{subfigure}
\usepackage{comment}
\usepackage{soul}
\usepackage{cancel}
\usepackage{dcolumn}

\usepackage{array}
\usepackage{ctable}
\usepackage{multirow}
\usepackage{siunitx}
\usepackage{longtable}
\usepackage{tabularx}
\usepackage{booktabs}
\usepackage{float}

\graphicspath{{Graphics/}}

\def\be{\begin{equation}}
\def\ee{\end{equation}}
\def\bea{\begin{eqnarray}}
\def\eea{\end{eqnarray}}

\def\prd{Phys. Rev. D}
\def\mnras{MNRAS}

\def\apj{ApJ}
\def\apjl{ApJ Lett.}

\def\aap{A\&A}

\def\jcap{JCAP}

\definecolor{vividviolet}{rgb}{0.62, 0.0, 1.0}
\definecolor{amaranth}{rgb}{0.9, 0.17, 0.31}
\definecolor{palatinateblue}{rgb}{0.15, 0.23, 0.89}
\definecolor{brightpink}{rgb}{1.0, 0.0, 0.5}
\definecolor{cornflowerblue}{rgb}{0.39, 0.58, 0.93}
\definecolor{deepcarminepink}{rgb}{0.94, 0.19, 0.22}
\definecolor{radicalred}{rgb}{1.0, 0.21, 0.37}

\hypersetup{ linktoc=all,
    colorlinks, linkcolor={palatinateblue},
    citecolor={brightpink}, urlcolor={amaranth}
}

\DeclareUnicodeCharacter{2212}{\ensuremath{-}}

\begin{document}

\title{Determining $H_0$  from distance sum rule combining gamma-ray bursts with observational Hubble data and strong gravitational lensing}

\author{Orlando Luongo}
\email{orlando.luongo@unicam.it}
\affiliation{Universit\`a di Camerino, Divisione di Fisica, Via Madonna delle carceri 9, 62032 Camerino, Italy.}
\affiliation{SUNY Polytechnic Institute, 13502 Utica, New York, USA.}
\affiliation{INAF, Osservatorio Astronomico di Brera, Milano, Italy.}
\affiliation{INFN, Sezione di Perugia, Perugia, 06123, Italy.}
\affiliation{Al-Farabi Kazakh National University, Al-Farabi av. 71, 050040 Almaty, Kazakhstan.}

\author{Marco Muccino}
\email{marco.muccino@lnf.infn.it}
\affiliation{Universit\`a di Camerino, Divisione di Fisica, Via Madonna delle carceri 9, 62032 Camerino, Italy.}
\affiliation{Al-Farabi Kazakh National University, Al-Farabi av. 71, 050040 Almaty, Kazakhstan.}
\affiliation{ICRANet, Piazza della Repubblica 10, 65122 Pescara, Italy.}

\begin{abstract}
Model-independent bounds on the Hubble constant $H_0$ are important to shed light on cosmological tensions.
We work out a model-independent analysis based on the sum rule, which is applied to late- and early-time data catalogs to determine $H_0$.
Through the model-independent B\'ezier interpolation of the observational Hubble data (OHD) and assuming a flat universe, we reconstruct the dimensionless distances of the sum rule and apply them to strong lensing data to derive constraints on $H_0$. Next, we extend this method to the high-redshift domain including, in other two separated analyses, gamma-ray burst (GRB) data sets from the well-established Amati and Combo correlations.
In all three analyses, our findings agree at $1\sigma$ level with the $H_0$ determined from type Ia supernovae (SNe Ia), and only at $2\sigma$ level with the measurement derived from the cosmic microwave background (CMB) radiation.
Our method evidences that the bounds on $H_0$ are significantly affected by strong lensing data, which favor the local measurement from SNe Ia. Including GRBs causes only a negligible decrease in the value of $H_0$. This may indicate that GRBs can be used to trace the expansion history and, in conjunction with CMB measurements, may heal the Hubble tension and accommodate to the flat $\Lambda$CDM paradigm purported by CMB data. 
\end{abstract}



\maketitle
\tableofcontents

\section{Introduction}

The Hubble constant $H_0$ defines current Universe's expansion rate and is one of the most crucial cosmological parameters to measure. It represents the zeroth term in any cosmographic expansion \cite{2012PhRvD..86l3516A,2017PDU....17...25A,2016IJGMM..1330002D} and, in fact, cosmology was primarily focused on determining $H_0$ and $q_0$, the deceleration parameter \citep{1970PhT....23b..34S}, albeit the need of extending the analysis to higher-orders has become increasingly evident recently, see e.g. Refs.~\citep{2013ApJ...769..133N,Luongo:2013rba}.

Despite its critical role in cosmology, current $H_0$ bounds differ significantly, highlighting a persistent and nowadays unresolved tension, possibly related to the emergence of new physics \citep{2024ApJ...975L..36H} or to some unexpected physical process and/or mechanism \citep{2020PhRvD.102b3518V}.

Nevertheless, the standard $\Lambda$CDM cosmological model, based on a bare cosmological constant $\Lambda$ that describes dark energy, is currently criticized, since recent results from \citet{2024arXiv240403002D} may favor a mildly evolving scalar field, even at the level of background cosmology, pushing up the Universe to accelerate through a repulsive effect over standard gravity \citep{2014PhRvD..90h4032L,Luongo:2015zaa}, even if several criticisms have been raised immediately after the release of their preliminary results, see e.g. Refs.~\cite{2024arXiv240408633C,2024A&A...690A..40L,2024arXiv240412068C}.

Using the $\Lambda$CDM model, bounds on $H_0$ seem to agree with the 
\textit{Planck} satellite ones, namely deeply in tension with respect to those found from SNe Ia\footnote{$H_0=(67.4\pm0.5)$~km~s$^{-1}$~Mpc$^{-1}$ from \citet{Planck2018} and $H_0=(74.03\pm 1.42)$~km~s$^{-1}$~Mpc$^{-1}$ from Riess \citep{Riess:2019cxk} yield a severe $4.4 \sigma$ confidence level tension.}. 

In view of such results, better understanding the $\Lambda$CDM drawbacks and inferring new strategies to constrain $H_0$ appear extremely needful in precision cosmology \citep{2022PDU....3601045C,2024arXiv240917019W,2024MNRAS.528L..20A,2022arXiv221102129C,2024PDU....4401464O,2022PhRvD.106d1301O,2015arXiv151207076L} and, consequently, fixing the spatial geometry density parameter $\Omega_k$, may lead to unexpected inconsistency, for example a higher lensing amplitude in the
cosmic microwave background (CMB) power spectra \citep{Planck2018}.

Accordingly, the Hubble constant can be determined from time delay measurements of gravitationally-lensed supernovae \citep{Refsdal:1964nw}, and such data set can be expanded including lensed quasars, since their brightness and variable nature can be used to  measure the so-called ``time-delay distance" 
\citep{Kelly:2014mwa,Goobar:2016uuf}. 
The latter represents a combination of three angular diameter distances among the
observer, lens, and source, but it suffers from the unsolved issue of a \emph{strong model dependence}\footnote{In other words, such measurements are jeopardized by the strong inconsistency of fixing the cosmological model \emph{a priori}, implying very different results, as due to the fact that the right cosmological model is still under scrutiny as its form appears quite unconstrained \citep{2023PhRvD.108j3519W}. 
}. Relaxing dark energy not to behave as a constant would, in fact, imply severe departures from measuring  $H_0$ and, in general, the role of spatial curvature influences significantly every background theory different from general relativity, see e.g. Refs.~\cite{2018GReGr..50...53C,Capozziello:2019cav}. 

Fully model-independent treatments to determine $H_0$ and/or $\Omega_k$ are thus still object of impressive efforts and, in  this respect, the so-called \emph{distance sum rule} offers an effective method to determine the spatial curvature and the Hubble constant without relying on a specific cosmological model \citep{Rasanen:2014mca}. 
Within this framework, significant efforts have been made in recent studies to constrain cosmic curvature \citep{Qi:2018atg,Wang:2019yob,Xia:2016dgk,Li:2018hyr,Zhou:2019vou}. These studies utilize precise measurements of the source-lens/observer distance ratio in strong gravitational lensing (SGL) systems \citep{Cao:2015qja,Chen:2018jcf}, as well as luminosity distances derived from various other distance indicators \citep{Qi:2018aio,Liao:2019hfl}.
However, this methodology is only weakly dependent on the Hubble constant, as 
$H_0$  enters not through a direct distance measurement but rather through a distance ratio \citep{cao2012constraints}. 

It is worth noting that for the effective implementation of the distance sum rule, utilizing distance probes at higher redshifts would be advantageous, as this would enrich the sample of SGL points with alternative catalogs.

Motivated by the above considerations, in this paper, we consider the sum rule in order to constrain the Hubble constant, including high redshift data points and  adopting a model-independent treatment to describe the Hubble rate at different redshifts. To do so, we employ gamma-ray bursts (GRB) as high-redshift distance indicators, adding the observational Hubble rate data (OHD), directly measured at different redshifts, together with data from SGL of quasars. In this respect, we single out two GRB correlations among the most promising ones, namely the $E_p$--$E_{iso}$ (or Amati) and $L_0$--$E_p$--$T$ (or Combo) correlations, while fixing the background through the use of the model-independent B\'ezier polynomials. The latter is a technique that approximates the Hubble rate up to the redshifts of our interest, avoiding \emph{de facto} to postulate a dark energy model \emph{a priori}, see Ref.~\cite{2019MNRAS.486L..46A}. In our treatment, we force a vanishing spatial curvature, to avoid model-dependence on the cosmic parameters in the distance determination. Accordingly, we perform three analyses: the first combining OHD and SGL catalogs, the second including the high-redshift GRB data from the $E_p$--$E_{iso}$ correlation, and finally replacing the latter with the $L_0$--$E_p$--$T$ correlation. Each analysis is based on the Markov Chain Monte Carlo (MCMC) method and the Metropolis-Hastings algorithm, run by modifying the free \textit{Wolfram Mathematica} code presented in Ref.~\cite{2019PhRvD..99d3516A}. Our findings indicate $H_0$ close to the local value inferred by Riess \cite{Riess:2019cxk}, in agreement with the expectations coming from lensed quasars. Accordingly, the Hubble tension is discussed, comparing our findings with \citet{Planck2018}. Implications for dark energy evolutions show that the standard cosmological background is well-approximated by a flat $\Lambda$CDM model, agreeing with the most recent developments \citep{2024A&A...690A..40L,2024arXiv240412068C} that criticize the preliminary results obtained by \citet{2024arXiv240403002D} that, instead, suggests a slightly-varying dark energy evolution. 

The paper is structured as follows. In Sec.~\ref{sec:2}, we introduce our methodology, based on the B\'ezier polynomial interpolation and the sum rule, GRB correlations, as well as OHD and lensed data catalogs. In Sec.~\ref{sec:3}, the three MCMC analyses and their corresponding results are shown. Finally, Sec.~\ref{sec:4} deals final developments, physical interpretation and perspectives of our work.

\section{Methods and cosmic data sets}
\label{sec:2}

Considering the validity of the cosmological principle\footnote{For a recent criticism on this, see e.g., Ref.~\cite{2022PhRvD.105j3510L}.}, the corresponding maximally symmetric spacetime well describing the Universe is the Friedmann-Robertson-Walker (FRW) metric. 
Accordingly, the comoving distance of a source at redshift $z_s$, with emission time $t_s$, as seen by an observer at redshift $z_l$, characterized by a time $t_l > t_s$ at which the observation is performed, is defined by
\begin{equation}
\label{comdis}
    d_M(z_l,z_s) = \frac{c}{H_0\sqrt{\Omega_k}}\sinh{\left[\int^{z_s}_{z_l} \frac{H_0\sqrt{\Omega_k}}{H(z^\prime)} dz^\prime\right]}\,,
\end{equation}
where $c$ is the speed of light. 

Conveniently, introducing the dimensionless comoving distance, $d=H_0 d_M/c$, and defining
\begin{subequations}
\label{positiones}
\begin{align}
&d_l\equiv d(0,z_l),\\
&d_s\equiv d(0,z_s),\\
&d_{ls}\equiv d(z_l,z_s),
\end{align}
\end{subequations}
from Eq.~\eqref{comdis}, we derive the well-known \emph{distance sum rule}, for a generic non-flat FRW Universe \citep{Rasanen:2014mca}
\begin{equation} 
\label{smr}
\frac{d_{ls}}{d_s}=\sqrt{1+\Omega_k d_l^2}-\frac{d_l}{d_s}\sqrt{1+\Omega_k
d_s^2}\,,
\end{equation}
that appears as a ratio among distances defined in Eqs.~\eqref{positiones}. 

Hence, knowing $d_l$ and $d_s$ enables constraints on $\Omega_k$ \citep{Qi:2018aio} and tests on the validity of the FRW metric
\citep{Qi:2018atg,cao2019direct}. 

Accordingly, Eq.~\eqref{smr} can be also written as
\begin{equation} 
\label{timedelay1}
\frac{d_{l}d_{s}}{d_{ls}}=\frac{1}{\sqrt{1/d_l^2+\Omega_k}-\sqrt{1/d_s^2+\Omega_k}}.
\end{equation}

Now, the difference with respect to Eq.~\eqref{smr} is that the ratio $d_{l}d_{s}/d_{ls}$ defines a (dimensionless) distance. 

If we measure $d_l$ and $d_s$ and we know the above ratio from an observable in terms of dimensional distances, then Eq.~\eqref{timedelay1} can also be used to constrain $H_0$.

\subsection{Dimensionless distances from OHD} 

The OHD catalog \citep{2024arXiv241104901L} consists of $N_O = 34$ measurements $H_k$ of the Hubble rate obtained considering by considering passively evolving galaxies at various redshifts $z$ as \emph{cosmic chronometers}. Assuming that in these galaxies all stars formed at the same time $\Delta t$, through their spectroscopic redshift measurements, the Hubble rate is derived as $H(z) \simeq -(1+z)^{-1}(\Delta z/\Delta t)$ \citep{2002ApJ...573...37J,2022LRR....25....6M}. 

Hence, the catalog can be interpolated by the well-established second-order B\'ezier curve \citep{2019MNRAS.486L..46A,2021MNRAS.503.4581L,2021MNRAS.501.3515M,2023MNRAS.518.2247L,2023MNRAS.523.4938M,2024JHEAp..42..178A,2024A&A...686A..30A,2024arXiv240802536A,2024arXiv241104901L} 
\begin{equation}
\label{bezier1}
\mathcal H(z) = \frac{\alpha_\star}{z_{\rm m}^2} \left[\alpha_0(z_{\rm m}-z)^2 + 2\alpha_1 z(z_{\rm m}-z) +\alpha_2 z^2\right]\,,
\end{equation}
where $\alpha_\star=100\,{\rm km\,s}^{-1}{\rm Mpc}^{-1}$ is a normalization, $\alpha_i$ are the B\'ezier coefficients, and $z_{\rm m}=1.965$ is the largest redshift of the OHD catalog. 
At $z=0$ from Eq.~\eqref{bezier1} we define the dimensionless Hubble constant as $h_0\equiv H_0/\alpha_\star\equiv\alpha_0$.

Although Eq.~\eqref{bezier1} is model-independent, the reconstructed $d(z)$ is independent from any cosmological parameters only if assuming a zero spatial curvature, i.e.,
\begin{equation}\label{callumdist}
    d(z)=\int^{z}_{0} \frac{\alpha_\star \alpha_0 dz'}{\mathcal H(z')}\,.
\end{equation}
\emph{In view of this assumption, 
Eqs.~\eqref{smr}--\eqref{timedelay1} will be considered with $\Omega_k\equiv0$ and, thus, the following analyses will focus only on the determination of the Hubble constant.}

\subsection{Constraints at high redshifts with GRBs}

Correlations have been proposed as tools to standardize GRB and probe the Universe up to $z\sim 8$--$9$ \citep{AmatiDellaValle2013,Yonetoku,ghirlanda,2021ApJ...908..181M,2021Galax...9...77L,2022MNRAS.512..439C}.
Below, we use two well-consolidated correlations. 
\begin{itemize}
\item[-] \emph{Amati or $E_p$--$E_{iso}$ correlation.}
This is a prompt emission correlation expressed as \citep{AmatiDellaValle2013}
\begin{equation}
\label{Amati_correlation}
    \log{E_p}=a+b\left(\log{E_{iso}}-52\right)\,,
\end{equation}
with intercept $a$, slope $b$ and dispersion $\sigma$. It correlates the rest-frame peak energy $E_p$ (in keV units) of the $\gamma$-ray energy spectrum with the isotropic radiated energy (in erg units), integrated over the total duration of the burst and defined as \citep{2021Galax...9...77L} 
\begin{equation}
\label{Eiso}
E_{iso} = 4\pi d_L^2(z)S_b(1+z)^{-1}\,,
\end{equation}
where $S_b$ (in erg~cm$^2$ units) is the observed bolometric fluence, in the rest-frame $1$--$10^4$~keV band. 
In particular, from Eqs.~\eqref{Amati_correlation}--\eqref{Eiso}, the luminosity distance $d_L(z)=c(1+z)d(z)/H_0$ can be expressed as
\begin{equation}
\label{dl_Amati}
\log{d_L} = \frac{1}{2}\left[\frac{\log{E_p}-a}{b} - \log{\left(\frac{S_b}{1+z}\right)}\right]+d_0^A\,,
\end{equation}
where $d_0^A=0.96$ includes all the constants and the conversion factor from cm to Mpc.
\item[-] \emph{Combo or $L_0$--$E_p$--$T$ correlation.}
This is a hybrid prompt--afterglow correlation \citep{2021ApJ...908..181M}
\begin{equation}
\label{Combo_correlation}
\log{L_0} = a + b \log{E_p} - \log{T}\,,
\end{equation}
with intercept $a$, slope $b$ and dispersion $\sigma$.
It correlates the prompt observable $E_p$ (in keV units) with the $0.3$--$10$~keV rest-frame X-ray afterglow luminosity $L_0$ (in erg~s$^{-1}$ units) and effective rest-frame duration $T$ (in seconds) of the plateau phase \citep{2021Galax...9...77L}. 
$L_0$ is related to the rest-frame $0.3$--$10$~keV plateau flux $F_0$ (in erg~cm$^2$~s$^{-1}$ units) through
\begin{equation}
\label{L0}
    L_0=4\pi d^2_L(z)F_0\,.
\end{equation}
From Eqs.~\eqref{Combo_correlation}--\eqref{L0}, the luminosity distance reads
\begin{equation}
\label{dl_Combo}
\log{d_L} = \frac{1}{2}\left[a + b\log{E_p} - \log{T} - \log{F_0}\right]+d_0^C\,,
\end{equation}
where $d_0^C=-25.04$ includes all the constants and the conversion factor from cm to Mpc.
\end{itemize}
Eqs.~\eqref{dl_Amati} and \eqref{dl_Combo} can be separately compared with Eq.~\eqref{callumdist} to extend the $d$--$z$ diagram to $z\sim8$--$9$, paying the prize of constraining also the correlation parameters ($a$, $b$ and $\sigma$ for each correlation). 

In our analyses, we consider the most updated and refined catalogs composed of $N_A=118$ sources for the $E_p$--$E_{iso}$ correlation \citep{2021JCAP...09..042K} and $N_C=174$ bursts for the $L_0$--$E_p$--$T$ correlation \citep{2021ApJ...908..181M}.

\subsection{Time-delay measurements from lensed quasars}

In SGL as the source and an early-type galaxy acting as the lens, the time delays $\Delta t_{ij}$ between lensed images $i$ and $j$ and the gravitational potential of the lensing galaxy $\Delta\phi_{ij}$ are related via the time-delay distance $D$
\citep{2021MNRAS.503.2179Q}
\begin{equation}
D = \frac{c \Delta t_{ij}}{\Delta
\phi_{ij}}\,.
\label{relation}
\end{equation}
Using the definition $D\equiv (1+z_l) d_{A,l} d_{A,s}/d_{A,ls}$, where $d_A(z)=cd(z)/[H_0(1+z)]$ is the angular distance, we can interpolate the time-delay distance \citep{2021MNRAS.503.2179Q}
\begin{equation}
\label{Ddt}
\mathcal D=\frac{c}{\alpha_\star \alpha_0}\frac{d_l d_s}{d_{ls}}.
\end{equation}
Combining Eqs.~\eqref{timedelay1} and \eqref{Ddt} and knowing $d_l$ and $d_s$ from OHD interpolation, the time-delay measurements of $\Delta t$
can contribute in determining the value of $H_0$ or $h_0\equiv\alpha_0$ without
involving \emph{a priori} cosmological models.

We consider $N_L=7$ quasar-galaxy SGL systems with precise determinations of $D$ used in Ref.~\cite{2021MNRAS.503.2179Q}.
The systems released by the H0LiCOW collaboration \citep{Wong:2019kwg} are B1608+656, RXJ1131-1231, HE 0435-1223, SDSS 1206+4332, WFI2033-4723 and PG 1115+080, whereas the one from the STRIDES collaboration \citep{shajib2019strides} is DES J0408-5354. All sources cover a range $z_s\in[0.654,2.375]$.

\section{MCMC analyses and results}
\label{sec:3}

\begin{table*}
\caption{We here report our best-fit results, given with $1$--$\sigma$ ($2$--$\sigma$) error bars and split into the three fitting procedures, based on our hierarchy choices. The first fit (MCMC~1) employs small redshift data only, from OHD and SGL catalogs. The second (MCMC~2) and third (MCMC~3), instead, involve GRB data from two GRB correlations, namely the Amati and Combo correlations, respectively.}
\centering
\footnotesize
\setlength{\tabcolsep}{0.6em}
\renewcommand{\arraystretch}{1.5}
\begin{tabular}{lcccccc}
\hline\hline
                                &
$\alpha_0\equiv h_0$            &   
$\alpha_1$                      &
$\alpha_2$                      &
$a$                             &
$b$                             &
$\sigma$                        \\
\hline   
MCMC~1                                      &
$0.764_{-0.046\,(0.071)}^{+0.034\,(0.062)}$ &
$0.867_{-0.166\,(0.271)}^{+0.168\,(0.259)}$ &
$2.172_{-0.283\,(0.448)}^{+0.295\,(0.471)}$ &
$-$                                         & 
$-$                                         &
$-$                                         \\
MCMC~2                                      &
$0.750_{-0.040\,(0.071)}^{+0.044\,(0.071)}$ &
$0.902_{-0.172\,(0.274)}^{+0.156\,(0.264)}$ &
$2.169_{-0.289\,(0.456)}^{+0.273\,(0.451)}$ &
$1.648_{-0.158\,(0.276)}^{+0.119\,(0.193)}$ & 
$0.855_{-0.089\,(0.139)}^{+0.108\,(0.191)}$ &
$0.331_{-0.048\,(0.072)}^{+0.066\,(0.114)}$ \\
MCMC~3                                      &
$0.752_{-0.041\,(0.066)}^{+0.040\,(0.066)}$ &
$0.888_{-0.161\,(0.258)}^{+0.162\,(0.265)}$ &
$2.295_{-0.283\,(0.439)}^{+0.266\,(0.443)}$ &
$49.613_{-0.344\,(0.556)}^{+0.370\,(0.598)}$ & 
$0.835_{-0.129\,(0.221)}^{+0.145\,(0.223)}$ &
$0.164_{-0.023\,(0.035)}^{+0.022\,(0.036)}$ \\ 
\hline
\end{tabular}
\label{tab:bestfit}
\end{table*}

We perform three analyses:
\begin{itemize}
\item[-] \emph{MCMC~1} that considers only low-redshift catalogs, i.e., OHD and SGL,
\item[-] \emph{MCMC~2} that, in addition to OHD and SGL data, includes also the $E_p$--$E_{iso}$ catalog, and 
\item[-] \emph{MCMC~3} that utilizes the $L_0$--$E_p$--$T$ catalog, instead of the $E_p$--$E_{iso}$ one.  
\end{itemize}

These analyses do not introduce a hierarchy among free coefficients or among data sets, but rather they are intended to show the differences in the numerical outcomes once late- and early-time catalogs of data points are included. 

To do so, we describe below each MCMC procedure, resulting with the corresponding fitting functions and bounds. 

\subsection{MCMC~1 outcomes}

To determine the dimensionless distances $d_s$ and $d_l$, we need to estimate the best-fit parameters $\alpha_0\equiv h_0$, $\alpha_1$ and $\alpha_2$. 
This is done by using the OHD catalog and maximizing the log-likelihood function 
\begin{equation}
    \ln\mathcal{L}_O = -\frac{1}{2} \sum^{N_O}_{k=1} \left\{ \ln(2\pi\sigma^2_{H_k}) + \left[\frac{H_k-\mathcal H(z_k)}{\sigma_{H_k}}\right]^2\right\},
\end{equation}
where $\sigma_{H_k}$ are the errors attached to $H_k$.

To strengthen the constrain on $\alpha_0\equiv h_0$, we employ the SGL data set.
In this case, the best-fit parameters are determined by maximizing the following log-likelihood function 
\begin{equation}
    \ln\mathcal{L}_L = -\frac{1}{2} \sum^{N_L}_{k=1} \left\{ \ln(2\pi\sigma^2_{D_k}) + \left[\frac{D_k-\mathcal D(z_k)}{\sigma_{D_k}}\right]^2\right\},
\end{equation}
where $\sigma_{D_k}$ are the errors attached to $D_k$.

Hence, joint constraints from OHD and SGL catalogs can be obtained by maximizing the combined log-likelihood function
\begin{equation}
    \ln\mathcal{L}_1 = \ln\mathcal L_O + \ln\mathcal L_L\,.
\end{equation}

\begin{figure}
\centering
\includegraphics[width=0.88\hsize,clip]{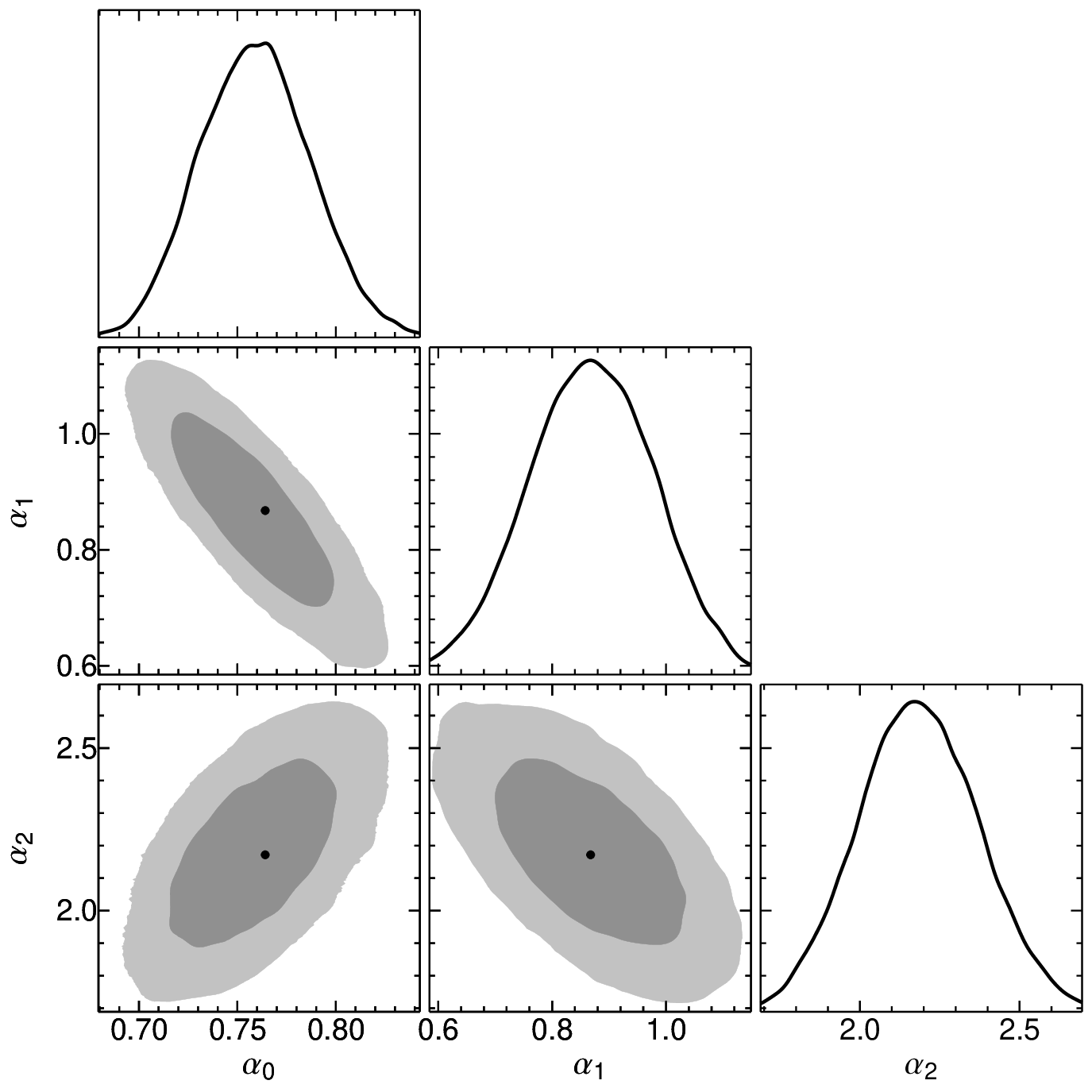}
\caption{MCMC~1 contour plots. Darker (lighter) areas display $1$--$\sigma$ ($2$--$\sigma$) confidence regions.}
\label{fig:cont1}
\end{figure}

The results are summarized in Table~\ref{tab:bestfit} and displayed in Fig.~\ref{fig:cont1}. 
We immediately notice that the inclusion of SGL data provides B\'ezier coefficients $\alpha_i$ with attached errors larger than those obtained by considering OHD only
\citep{2019MNRAS.486L..46A,2021MNRAS.503.4581L,2021MNRAS.501.3515M,2023MNRAS.518.2247L,2023MNRAS.523.4938M,2024JHEAp..42..178A,2024A&A...686A..30A,2024arXiv240802536A,2024arXiv241104901L}.
Moreover, while $\alpha_1$ and $\alpha_2$ are in line with those obtained by considering OHD only, the lowest order coefficient $\alpha_0\equiv h_0$ has shifted from the usual value $\alpha_0\simeq 0.68$--$0.69$ \citep{2019MNRAS.486L..46A,2021MNRAS.503.4581L,2021MNRAS.501.3515M,2023MNRAS.518.2247L,2023MNRAS.523.4938M,2024JHEAp..42..178A,2024A&A...686A..30A,2024arXiv240802536A,2024arXiv241104901L} to a higher value $\alpha_0\simeq 0.76$ which is consistent with the results based on SGL data applied to the distance sum rule \citep{2021MNRAS.503.2179Q}.

\subsection{MCMC~2 outcomes}

Now, we include the GRBs of the $E_p$--$E_{iso}$ data set to check to what extent the B\'ezier coefficients, in particular $\alpha_0$, are affected by extending the $d$--$z$ diagram to $z\sim8$--$9$.
Worth noticing, the inclusion of this data set introduces additional free parameters ($a$, $b$ and $\sigma$) to be evaluated, possibly deteriorating the best-fit accuracy. 

The log-likelihood for the GRBs of the $E_p$--$E_{iso}$ (or Amati) correlation is given by
\begin{align}
\nonumber
\ln\mathcal{L}_A = &-\frac{1}{2} \sum^{N_A}_{k=1}  \left[\frac{\log{(\alpha_\star^{-1}\alpha_0^{-1}c d_k)} -\log{d_{L,k}}}{\sigma_{log{d_{L,k}}}}\right]^2\\ 
&-\frac{1}{2} \sum^{N_A}_{k=1} \ln(2\pi\sigma^2_{log{d_{L,k}}})\,,
\end{align}
where $d_k$ is defined in Eq.~\eqref{callumdist} and $\log{d_{L,k}}$ in Eq.~\eqref{dl_Amati}.

The best-fit B\'ezier coefficients $\alpha_i$ and the correlation parameters $a$, $b$ and $\sigma$ are, thus, obtained by maximizing the combined log-likelihood function
\begin{equation}
    \ln\mathcal{L}_2 = \ln\mathcal L_1 + \ln\mathcal L_A\,.
\end{equation}

\begin{figure*}
\centering
\includegraphics[width=0.78\hsize,clip]{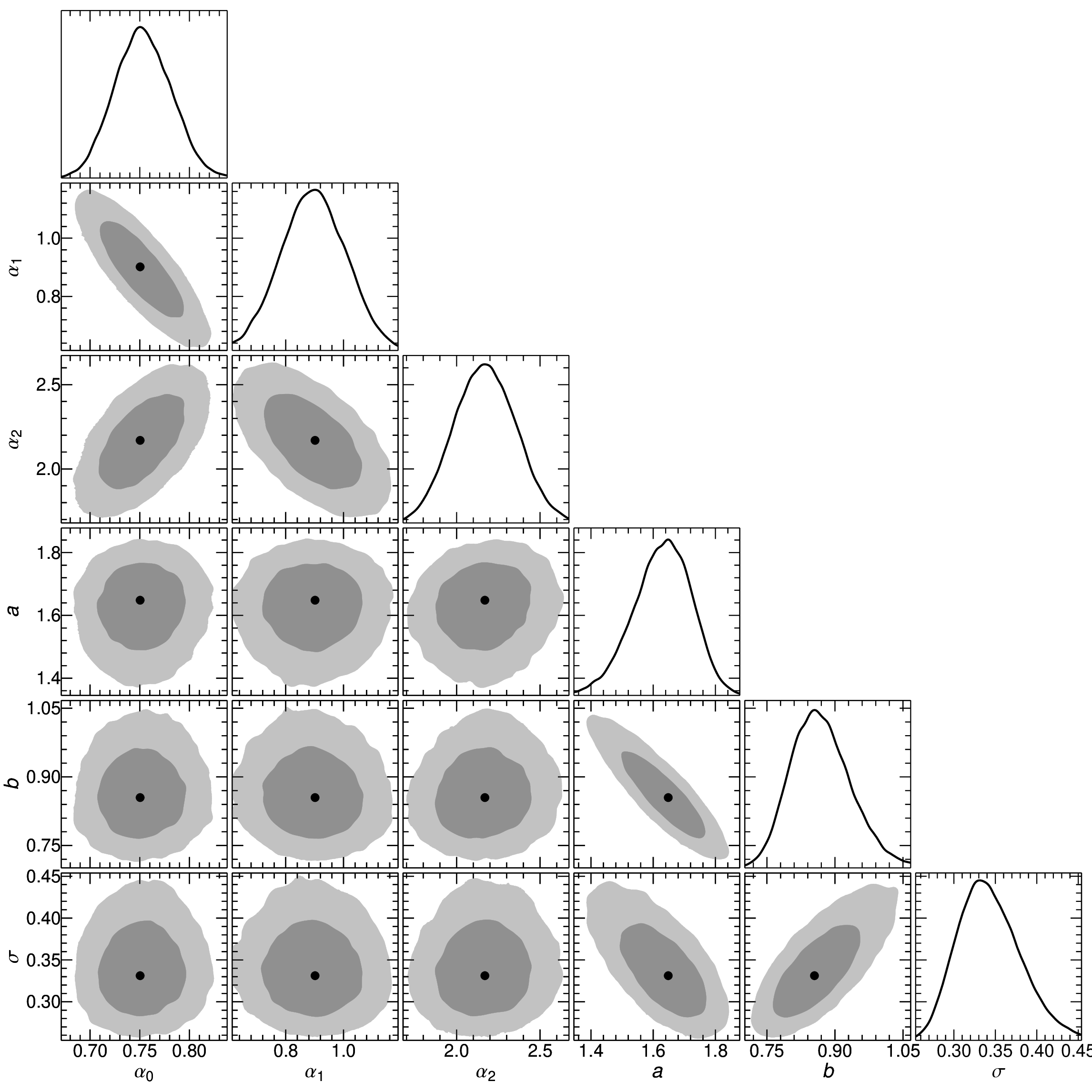}
\caption{MCMC~2 contour plots. Darker (lighter) areas display $1$--$\sigma$ ($2$--$\sigma$) confidence regions.}
\label{fig:cont2}
\end{figure*}

The results are summarized in Table~\ref{tab:bestfit} and displayed in Fig.~\ref{fig:cont2}.
With respect to MCMC~1, the inclusion of GRB data neither changes significantly the B\'ezier best-fit coefficients $\alpha_i$, nor increases the attached errors.
The value of $\alpha_0\equiv h_0$ has slightly decreased with respect to the MCMC~1 analysis, but it is still high ($\alpha_0\simeq 0.75$) and significantly influenced by SGL data \citep{2021MNRAS.503.2179Q}.

Regarding the $E_p$--$E_{iso}$ correlation parameters, their values are in line with those obtained by calibrating this correlation with the OHD catalog \citep{2019MNRAS.486L..46A,2021MNRAS.503.4581L,2021MNRAS.501.3515M,2023MNRAS.518.2247L,2023MNRAS.523.4938M,2024JHEAp..42..178A,2024A&A...686A..30A,2024arXiv240802536A}.

\subsection{MCMC~3 outcomes}

We now consider the GRBs from the $L_0$--$E_p$--$T$ correlation. The general implications of adding a GRB catalog, that requires to fit additional parameters, still hold for this case. 

Like the previous case, the log-likelihood for the GRBs of the $L_0$--$E_p$--$T$ (or Combo) correlation is given by
\begin{align}
\nonumber
\ln\mathcal{L}_C = &-\frac{1}{2} \sum^{N_C}_{k=1}  \left[\frac{\log{(\alpha_\star^{-1}\alpha_0^{-1}c d_k)} -\log{d_{L,k}}}{\sigma_{log{d_{L,k}}}}\right]^2\\ 
&-\frac{1}{2} \sum^{N_C}_{k=1} \ln(2\pi\sigma^2_{log{d_{L,k}}})\,,
\end{align}
where $d_k$ is again defined by Eq.~\eqref{callumdist}, but $\log{d_{L,k}}$ is now given by Eq.~\eqref{dl_Combo}.

\begin{figure*}
\centering
\includegraphics[width=0.78\hsize,clip]{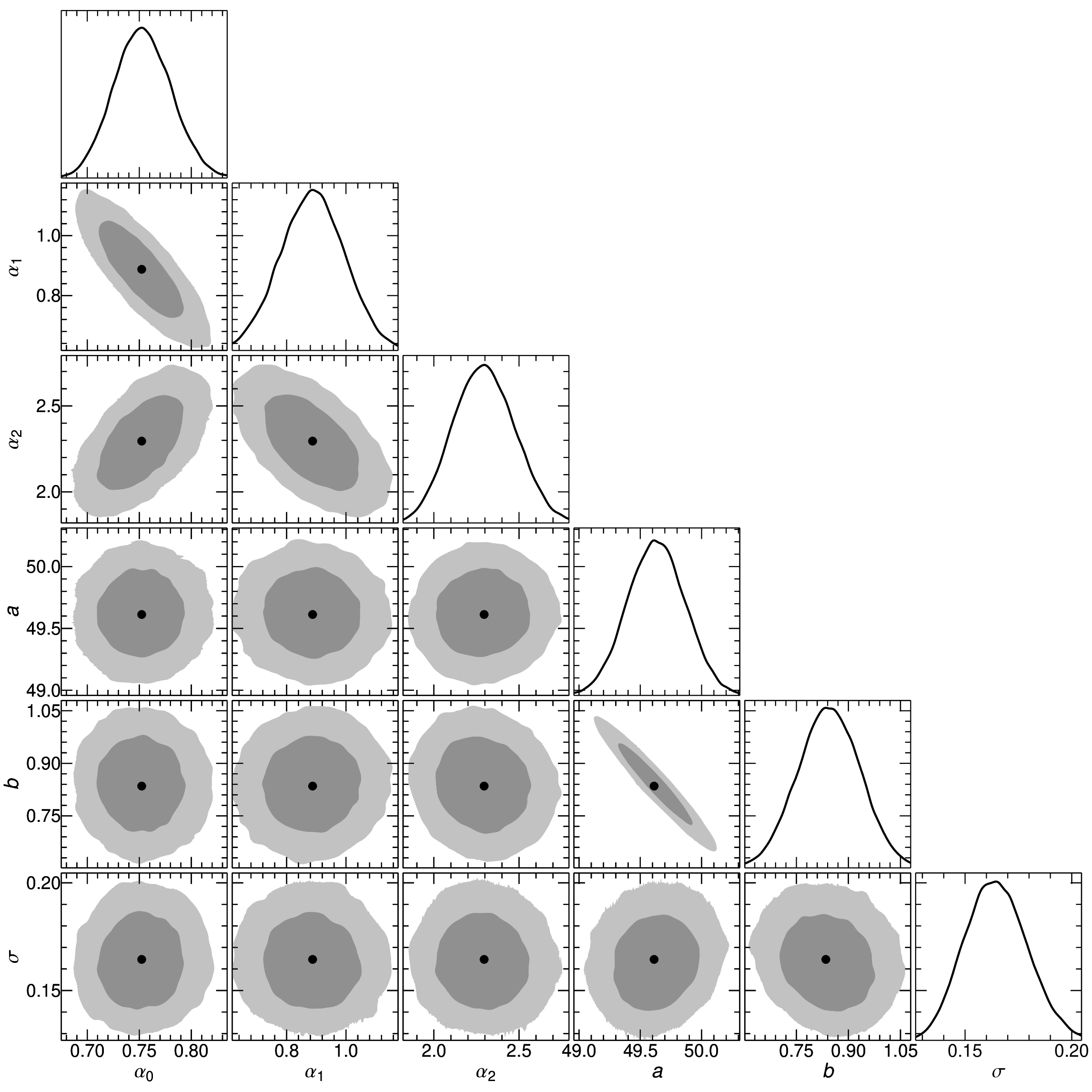}
\caption{MCMC~3 contour plots. Darker (lighter) areas display $1$--$\sigma$ ($2$--$\sigma$) confidence regions.}
\label{fig:cont3}
\end{figure*}

In this case, $\alpha_i$, $a$, $b$ and $\sigma$ are obtained by maximizing the combined log-likelihood function
\begin{equation}
    \ln\mathcal{L}_3 = \ln\mathcal L_1 + \ln\mathcal L_C\,.
\end{equation}
and the corresponding results are summarized in Table~\ref{tab:bestfit} and displayed in Fig.~\ref{fig:cont3}.

Changing GRB data set leads to results similar to the ones of the MCMC~2 analysis, both in terms of B\'ezier coefficients and correlation parameters.
Basically, the most noteworthy conclusions from the MCMC~3 analysis are that (a) the value of $\alpha_0\equiv h_0$ is, again, significantly influenced by SGL data \citep{2021MNRAS.503.2179Q}, and (b) the $L_0$--$E_p$--$T$ correlation is characterized by a very small dispersion, when compared to the $E_p$--$E_{iso}$.

\section{Final outlooks}
\label{sec:4}

Within model-independent reconstruction techniques, we here focused on determining $H_0$ bounds in order to  better understand the cosmological tension by reconciling late- and early-time data. 
To do so, we worked out a model-independent analysis involving the so-called \emph{sum rule}, for which we selected  late- and early-time data catalogs to estimate the universe rate today.

More precisely, we model-independently interpolated the most recent OHD catalog via B\'ezier second order function. 
In this respect, the B\'ezier polynomials are able to frame the universe dynamics similarly to splines, i.e., constructing a test function that is not based on the sum of cosmic fluids, as it happens for dark energy models. 
Further, to get model-independent determinations of the dimensionless distance from OHD and reconcile observations made at our time with the ones at high redshifts from  GRBs, we assumed the universe to be flat, i.e., $\Omega_k=0$. 

We performed three analyses. 
In the first one named MCMC~1, we employed only low-redshift catalogs, namely OHD, to reconstruct the dimensionless distances, combine them into the distance sum rule, and apply the latter to SGL data. 
The second and third analyses, respectively MCMC~2 and 3, extend our approach to the high-redshift regime incorporating, within two separate analyses, two GRB correlation data sets, namely the $E_p-E_{iso}$ correlation for MCMC~2 and the $L_0$--$E_p$--$T$ correlation for MCMC~3.

In all the aforementioned cases, we performed a MCMC analysis based on the Metropolis-Hastings algorithm. We derived constraints on $H_0$ that are consistent at $1\sigma$ level with the local measurement of $H_0$ obtained from SNe Ia \cite{Riess:2019cxk}, while quite unexpectedly the agreement is only at the $2\sigma$ level with the measurement based on the CMB radiation \citep{Planck2018}.

Our findings remarked the significant influence of SGL data, that agree more closely with the local supernova measurements on $H_0$. 
Moreover, our results showed that the inclusion of GRB data furnishes a minimal impact on determining $H_0$, likely indicating that GRBs could serve as effective probes of the expansion history and, at the same time, that it would be possible that the Hubble tension may be due to the need of reconciling  early- and late-time data rather than indicating new physics\footnote{Usually, the inclusion of GRBs may lead to unexpectedly large determinations of mass \citep{2021ApJ...908..181M}. Nevertheless, their use to constrain background cosmology may provide misleading results, as due to the calibration procedure \citep{2021JCAP...09..042K}. Accordingly, the combination of GRBs with low-redshift data catalogs may refine the analyses, suggesting possible deviations from the standard $\Lambda$CDM model \citep{2020A&A...641A.174L}. }. 
Moreover, combined with CMB radiation data, GRBs might help alleviating the Hubble tension while remaining compatible with the flat $\Lambda$CDM paradigm. 

Future works will be devoted to explore  these avenues, i.e., the influence of SGL  data on cosmic probes, the inclusion of high-redshift data inferred from GRBs and, possibly, testing again the dark energy models adopting generic lensing data and the sum rule, verifying how and whether the current expectations, obtained from international collaborations, such as DESI, Planck and the future Euclid missions, may be compared once including such data catalogs and the use of the cosmic distance sum rule.

\section*{Acknowledgements}
The authors express their gratitude for discussions to Deepak Jain about the topic of model-independent reconstructions in cosmology. OL expresses his gratitude to Sunny Vagnozzi for interactions in the field of cosmology and Hubble tension, during his stay at the University of Camerino. The authors are grateful to Giovanni Delle Monache for debates on the topic of statistics. 


\begin{thebibliography}{65}%
\makeatletter
\providecommand \@ifxundefined [1]{%
 \@ifx{#1\undefined}
}%
\providecommand \@ifnum [1]{%
 \ifnum #1\expandafter \@firstoftwo
 \else \expandafter \@secondoftwo
 \fi
}%
\providecommand \@ifx [1]{%
 \ifx #1\expandafter \@firstoftwo
 \else \expandafter \@secondoftwo
 \fi
}%
\providecommand \natexlab [1]{#1}%
\providecommand \enquote  [1]{``#1''}%
\providecommand \bibnamefont  [1]{#1}%
\providecommand \bibfnamefont [1]{#1}%
\providecommand \citenamefont [1]{#1}%
\providecommand \href@noop [0]{\@secondoftwo}%
\providecommand \href [0]{\begingroup \@sanitize@url \@href}%
\providecommand \@href[1]{\@@startlink{#1}\@@href}%
\providecommand \@@href[1]{\endgroup#1\@@endlink}%
\providecommand \@sanitize@url [0]{\catcode `\\12\catcode `\$12\catcode `\&12\catcode `\#12\catcode `\^12\catcode `\_12\catcode `\%12\relax}%
\providecommand \@@startlink[1]{}%
\providecommand \@@endlink[0]{}%
\providecommand \url  [0]{\begingroup\@sanitize@url \@url }%
\providecommand \@url [1]{\endgroup\@href {#1}{\urlprefix }}%
\providecommand \urlprefix  [0]{URL }%
\providecommand \Eprint [0]{\href }%
\providecommand \doibase [0]{http://dx.doi.org/}%
\providecommand \selectlanguage [0]{\@gobble}%
\providecommand \bibinfo  [0]{\@secondoftwo}%
\providecommand \bibfield  [0]{\@secondoftwo}%
\providecommand \translation [1]{[#1]}%
\providecommand \BibitemOpen [0]{}%
\providecommand \bibitemStop [0]{}%
\providecommand \bibitemNoStop [0]{.\EOS\space}%
\providecommand \EOS [0]{\spacefactor3000\relax}%
\providecommand \BibitemShut  [1]{\csname bibitem#1\endcsname}%
\let\auto@bib@innerbib\@empty
\bibitem [{\citenamefont {{Aviles}}\ \emph {et~al.}(2012)\citenamefont {{Aviles}}, \citenamefont {{Gruber}}, \citenamefont {{Luongo}},\ and\ \citenamefont {{Quevedo}}}]{2012PhRvD..86l3516A}%
  \BibitemOpen
  \bibfield  {author} {\bibinfo {author} {\bibfnamefont {A.}~\bibnamefont {{Aviles}}}, \bibinfo {author} {\bibfnamefont {C.}~\bibnamefont {{Gruber}}}, \bibinfo {author} {\bibfnamefont {O.}~\bibnamefont {{Luongo}}}, \ and\ \bibinfo {author} {\bibfnamefont {H.}~\bibnamefont {{Quevedo}}},\ }\href {\doibase 10.1103/PhysRevD.86.123516} {\bibfield  {journal} {\bibinfo  {journal} {\prd}\ }\textbf {\bibinfo {volume} {86}},\ \bibinfo {eid} {123516} (\bibinfo {year} {2012})},\ \Eprint {http://arxiv.org/abs/1204.2007} {arXiv:1204.2007 [astro-ph.CO]} \BibitemShut {NoStop}%
\bibitem [{\citenamefont {{Aviles}}\ \emph {et~al.}(2017)\citenamefont {{Aviles}}, \citenamefont {{Klapp}},\ and\ \citenamefont {{Luongo}}}]{2017PDU....17...25A}%
  \BibitemOpen
  \bibfield  {author} {\bibinfo {author} {\bibfnamefont {A.}~\bibnamefont {{Aviles}}}, \bibinfo {author} {\bibfnamefont {J.}~\bibnamefont {{Klapp}}}, \ and\ \bibinfo {author} {\bibfnamefont {O.}~\bibnamefont {{Luongo}}},\ }\href {\doibase 10.1016/j.dark.2017.07.002} {\bibfield  {journal} {\bibinfo  {journal} {Physics of the Dark Universe}\ }\textbf {\bibinfo {volume} {17}},\ \bibinfo {pages} {25} (\bibinfo {year} {2017})},\ \Eprint {http://arxiv.org/abs/1606.09195} {arXiv:1606.09195 [astro-ph.CO]} \BibitemShut {NoStop}%
\bibitem [{\citenamefont {{Dunsby}}\ and\ \citenamefont {{Luongo}}(2016)}]{2016IJGMM..1330002D}%
  \BibitemOpen
  \bibfield  {author} {\bibinfo {author} {\bibfnamefont {P.~K.~S.}\ \bibnamefont {{Dunsby}}}\ and\ \bibinfo {author} {\bibfnamefont {O.}~\bibnamefont {{Luongo}}},\ }\href {\doibase 10.1142/S0219887816300026} {\bibfield  {journal} {\bibinfo  {journal} {International Journal of Geometric Methods in Modern Physics}\ }\textbf {\bibinfo {volume} {13}},\ \bibinfo {eid} {1630002-606} (\bibinfo {year} {2016})},\ \Eprint {http://arxiv.org/abs/1511.06532} {arXiv:1511.06532 [gr-qc]} \BibitemShut {NoStop}%
\bibitem [{\citenamefont {{Sandage}}(1970)}]{1970PhT....23b..34S}%
  \BibitemOpen
  \bibfield  {author} {\bibinfo {author} {\bibfnamefont {A.~R.}\ \bibnamefont {{Sandage}}},\ }\href {\doibase 10.1063/1.3021960} {\bibfield  {journal} {\bibinfo  {journal} {Physics Today}\ }\textbf {\bibinfo {volume} {23}},\ \bibinfo {pages} {34} (\bibinfo {year} {1970})}\BibitemShut {NoStop}%
\bibitem [{\citenamefont {{Neben}}\ and\ \citenamefont {{Turner}}(2013)}]{2013ApJ...769..133N}%
  \BibitemOpen
  \bibfield  {author} {\bibinfo {author} {\bibfnamefont {A.~R.}\ \bibnamefont {{Neben}}}\ and\ \bibinfo {author} {\bibfnamefont {M.~S.}\ \bibnamefont {{Turner}}},\ }\href {\doibase 10.1088/0004-637X/769/2/133} {\bibfield  {journal} {\bibinfo  {journal} {\apj}\ }\textbf {\bibinfo {volume} {769}},\ \bibinfo {eid} {133} (\bibinfo {year} {2013})},\ \Eprint {http://arxiv.org/abs/1209.0480} {arXiv:1209.0480 [astro-ph.CO]} \BibitemShut {NoStop}%
\bibitem [{\citenamefont {Luongo}(2013)}]{Luongo:2013rba}%
  \BibitemOpen
  \bibfield  {author} {\bibinfo {author} {\bibfnamefont {O.}~\bibnamefont {Luongo}},\ }\href {\doibase 10.1142/S0217732313500806} {\bibfield  {journal} {\bibinfo  {journal} {Mod. Phys. Lett. A}\ }\textbf {\bibinfo {volume} {28}},\ \bibinfo {pages} {1350080} (\bibinfo {year} {2013})}\BibitemShut {NoStop}%
\bibitem [{\citenamefont {{Hu}}\ \emph {et~al.}(2024)\citenamefont {{Hu}}, \citenamefont {{Jia}}, \citenamefont {{Hu}},\ and\ \citenamefont {{Wang}}}]{2024ApJ...975L..36H}%
  \BibitemOpen
  \bibfield  {author} {\bibinfo {author} {\bibfnamefont {J.~P.}\ \bibnamefont {{Hu}}}, \bibinfo {author} {\bibfnamefont {X.~D.}\ \bibnamefont {{Jia}}}, \bibinfo {author} {\bibfnamefont {J.}~\bibnamefont {{Hu}}}, \ and\ \bibinfo {author} {\bibfnamefont {F.~Y.}\ \bibnamefont {{Wang}}},\ }\href {\doibase 10.3847/2041-8213/ad85cf} {\bibfield  {journal} {\bibinfo  {journal} {\apjl}\ }\textbf {\bibinfo {volume} {975}},\ \bibinfo {eid} {L36} (\bibinfo {year} {2024})},\ \Eprint {http://arxiv.org/abs/2410.06450} {arXiv:2410.06450 [astro-ph.CO]} \BibitemShut {NoStop}%
\bibitem [{\citenamefont {{Vagnozzi}}(2020)}]{2020PhRvD.102b3518V}%
  \BibitemOpen
  \bibfield  {author} {\bibinfo {author} {\bibfnamefont {S.}~\bibnamefont {{Vagnozzi}}},\ }\href {\doibase 10.1103/PhysRevD.102.023518} {\bibfield  {journal} {\bibinfo  {journal} {\prd}\ }\textbf {\bibinfo {volume} {102}},\ \bibinfo {eid} {023518} (\bibinfo {year} {2020})},\ \Eprint {http://arxiv.org/abs/1907.07569} {arXiv:1907.07569 [astro-ph.CO]} \BibitemShut {NoStop}%
\bibitem [{\citenamefont {{DESI Collaboration}}(2024)}]{2024arXiv240403002D}%
  \BibitemOpen
  \bibfield  {author} {\bibinfo {author} {\bibnamefont {{DESI Collaboration}}},\ }\href {\doibase 10.48550/arXiv.2404.03002} {\bibfield  {journal} {\bibinfo  {journal} {arXiv e-prints}\ ,\ \bibinfo {eid} {arXiv:2404.03002}} (\bibinfo {year} {2024})},\ \Eprint {http://arxiv.org/abs/2404.03002} {arXiv:2404.03002 [astro-ph.CO]} \BibitemShut {NoStop}%
\bibitem [{\citenamefont {{Luongo}}\ and\ \citenamefont {{Quevedo}}(2014)}]{2014PhRvD..90h4032L}%
  \BibitemOpen
  \bibfield  {author} {\bibinfo {author} {\bibfnamefont {O.}~\bibnamefont {{Luongo}}}\ and\ \bibinfo {author} {\bibfnamefont {H.}~\bibnamefont {{Quevedo}}},\ }\href {\doibase 10.1103/PhysRevD.90.084032} {\bibfield  {journal} {\bibinfo  {journal} {\prd}\ }\textbf {\bibinfo {volume} {90}},\ \bibinfo {eid} {084032} (\bibinfo {year} {2014})},\ \Eprint {http://arxiv.org/abs/1407.1530} {arXiv:1407.1530 [gr-qc]} \BibitemShut {NoStop}%
\bibitem [{\citenamefont {Luongo}\ and\ \citenamefont {Quevedo}(2018)}]{Luongo:2015zaa}%
  \BibitemOpen
  \bibfield  {author} {\bibinfo {author} {\bibfnamefont {O.}~\bibnamefont {Luongo}}\ and\ \bibinfo {author} {\bibfnamefont {H.}~\bibnamefont {Quevedo}},\ }\href {\doibase 10.1007/s10701-017-0125-0} {\bibfield  {journal} {\bibinfo  {journal} {Found. Phys.}\ }\textbf {\bibinfo {volume} {48}},\ \bibinfo {pages} {17} (\bibinfo {year} {2018})},\ \Eprint {http://arxiv.org/abs/1507.06446} {arXiv:1507.06446 [gr-qc]} \BibitemShut {NoStop}%
\bibitem [{\citenamefont {{Colg{\'a}in}}\ \emph {et~al.}(2024)\citenamefont {{Colg{\'a}in}}, \citenamefont {{Dainotti}}, \citenamefont {{Capozziello}}, \citenamefont {{Pourojaghi}}, \citenamefont {{Sheikh-Jabbari}},\ and\ \citenamefont {{Stojkovic}}}]{2024arXiv240408633C}%
  \BibitemOpen
  \bibfield  {author} {\bibinfo {author} {\bibfnamefont {E.~{\'O}.}\ \bibnamefont {{Colg{\'a}in}}}, \bibinfo {author} {\bibfnamefont {M.~G.}\ \bibnamefont {{Dainotti}}}, \bibinfo {author} {\bibfnamefont {S.}~\bibnamefont {{Capozziello}}}, \bibinfo {author} {\bibfnamefont {S.}~\bibnamefont {{Pourojaghi}}}, \bibinfo {author} {\bibfnamefont {M.~M.}\ \bibnamefont {{Sheikh-Jabbari}}}, \ and\ \bibinfo {author} {\bibfnamefont {D.}~\bibnamefont {{Stojkovic}}},\ }\href {\doibase 10.48550/arXiv.2404.08633} {\bibfield  {journal} {\bibinfo  {journal} {arXiv e-prints}\ ,\ \bibinfo {eid} {arXiv:2404.08633}} (\bibinfo {year} {2024})},\ \Eprint {http://arxiv.org/abs/2404.08633} {arXiv:2404.08633 [astro-ph.CO]} \BibitemShut {NoStop}%
\bibitem [{\citenamefont {{Luongo}}\ and\ \citenamefont {{Muccino}}(2024{\natexlab{a}})}]{2024A&A...690A..40L}%
  \BibitemOpen
  \bibfield  {author} {\bibinfo {author} {\bibfnamefont {O.}~\bibnamefont {{Luongo}}}\ and\ \bibinfo {author} {\bibfnamefont {M.}~\bibnamefont {{Muccino}}},\ }\href {\doibase 10.1051/0004-6361/202450512} {\bibfield  {journal} {\bibinfo  {journal} {\aap}\ }\textbf {\bibinfo {volume} {690}},\ \bibinfo {eid} {A40} (\bibinfo {year} {2024}{\natexlab{a}})},\ \Eprint {http://arxiv.org/abs/2404.07070} {arXiv:2404.07070 [astro-ph.CO]} \BibitemShut {NoStop}%
\bibitem [{\citenamefont {{Carloni}}\ \emph {et~al.}(2024)\citenamefont {{Carloni}}, \citenamefont {{Luongo}},\ and\ \citenamefont {{Muccino}}}]{2024arXiv240412068C}%
  \BibitemOpen
  \bibfield  {author} {\bibinfo {author} {\bibfnamefont {Y.}~\bibnamefont {{Carloni}}}, \bibinfo {author} {\bibfnamefont {O.}~\bibnamefont {{Luongo}}}, \ and\ \bibinfo {author} {\bibfnamefont {M.}~\bibnamefont {{Muccino}}},\ }\href {\doibase 10.48550/arXiv.2404.12068} {\bibfield  {journal} {\bibinfo  {journal} {arXiv e-prints}\ ,\ \bibinfo {eid} {arXiv:2404.12068}} (\bibinfo {year} {2024})},\ \Eprint {http://arxiv.org/abs/2404.12068} {arXiv:2404.12068 [astro-ph.CO]} \BibitemShut {NoStop}%
\bibitem [{\citenamefont {{Planck Collaboration}}(2020)}]{Planck2018}%
  \BibitemOpen
  \bibfield  {author} {\bibinfo {author} {\bibnamefont {{Planck Collaboration}}},\ }\href {\doibase 10.1051/0004-6361/201833910} {\bibfield  {journal} {\bibinfo  {journal} {A\&A}\ }\textbf {\bibinfo {volume} {641}},\ \bibinfo {eid} {A6} (\bibinfo {year} {2020})},\ \Eprint {http://arxiv.org/abs/1807.06209} {arXiv:1807.06209 [astro-ph.CO]} \BibitemShut {NoStop}%
\bibitem [{\citenamefont {Riess}\ \emph {et~al.}(2019)\citenamefont {Riess}, \citenamefont {Casertano}, \citenamefont {Yuan}, \citenamefont {Macri},\ and\ \citenamefont {Scolnic}}]{Riess:2019cxk}%
  \BibitemOpen
  \bibfield  {author} {\bibinfo {author} {\bibfnamefont {A.~G.}\ \bibnamefont {Riess}}, \bibinfo {author} {\bibfnamefont {S.}~\bibnamefont {Casertano}}, \bibinfo {author} {\bibfnamefont {W.}~\bibnamefont {Yuan}}, \bibinfo {author} {\bibfnamefont {L.~M.}\ \bibnamefont {Macri}}, \ and\ \bibinfo {author} {\bibfnamefont {D.}~\bibnamefont {Scolnic}},\ }\href {\doibase 10.3847/1538-4357/ab1422} {\bibfield  {journal} {\bibinfo  {journal} {Astrophys. J.}\ }\textbf {\bibinfo {volume} {876}},\ \bibinfo {pages} {85} (\bibinfo {year} {2019})},\ \Eprint {http://arxiv.org/abs/1903.07603} {arXiv:1903.07603 [astro-ph.CO]} \BibitemShut {NoStop}%
\bibitem [{\citenamefont {{Capozziello}}\ \emph {et~al.}(2022)\citenamefont {{Capozziello}}, \citenamefont {{D'Agostino}},\ and\ \citenamefont {{Luongo}}}]{2022PDU....3601045C}%
  \BibitemOpen
  \bibfield  {author} {\bibinfo {author} {\bibfnamefont {S.}~\bibnamefont {{Capozziello}}}, \bibinfo {author} {\bibfnamefont {R.}~\bibnamefont {{D'Agostino}}}, \ and\ \bibinfo {author} {\bibfnamefont {O.}~\bibnamefont {{Luongo}}},\ }\href {\doibase 10.1016/j.dark.2022.101045} {\bibfield  {journal} {\bibinfo  {journal} {Physics of the Dark Universe}\ }\textbf {\bibinfo {volume} {36}},\ \bibinfo {eid} {101045} (\bibinfo {year} {2022})},\ \Eprint {http://arxiv.org/abs/2202.03300} {arXiv:2202.03300 [astro-ph.CO]} \BibitemShut {NoStop}%
\bibitem [{\citenamefont {{Wolf}}\ \emph {et~al.}(2024)\citenamefont {{Wolf}}, \citenamefont {{Ferreira}},\ and\ \citenamefont {{Garc{\'\i}a-Garc{\'\i}a}}}]{2024arXiv240917019W}%
  \BibitemOpen
  \bibfield  {author} {\bibinfo {author} {\bibfnamefont {W.~J.}\ \bibnamefont {{Wolf}}}, \bibinfo {author} {\bibfnamefont {P.~G.}\ \bibnamefont {{Ferreira}}}, \ and\ \bibinfo {author} {\bibfnamefont {C.}~\bibnamefont {{Garc{\'\i}a-Garc{\'\i}a}}},\ }\href {\doibase 10.48550/arXiv.2409.17019} {\bibfield  {journal} {\bibinfo  {journal} {arXiv e-prints}\ ,\ \bibinfo {eid} {arXiv:2409.17019}} (\bibinfo {year} {2024})},\ \Eprint {http://arxiv.org/abs/2409.17019} {arXiv:2409.17019 [astro-ph.CO]} \BibitemShut {NoStop}%
\bibitem [{\citenamefont {{Adil}}\ \emph {et~al.}(2024)\citenamefont {{Adil}}, \citenamefont {{Akarsu}}, \citenamefont {{Malekjani}}, \citenamefont {{{\'O} Colg{\'a}in}}, \citenamefont {{Pourojaghi}}, \citenamefont {{Sen}},\ and\ \citenamefont {{Sheikh-Jabbari}}}]{2024MNRAS.528L..20A}%
  \BibitemOpen
  \bibfield  {author} {\bibinfo {author} {\bibfnamefont {S.~A.}\ \bibnamefont {{Adil}}}, \bibinfo {author} {\bibfnamefont {{\"O}.}~\bibnamefont {{Akarsu}}}, \bibinfo {author} {\bibfnamefont {M.}~\bibnamefont {{Malekjani}}}, \bibinfo {author} {\bibfnamefont {E.}~\bibnamefont {{{\'O} Colg{\'a}in}}}, \bibinfo {author} {\bibfnamefont {S.}~\bibnamefont {{Pourojaghi}}}, \bibinfo {author} {\bibfnamefont {A.~A.}\ \bibnamefont {{Sen}}}, \ and\ \bibinfo {author} {\bibfnamefont {M.~M.}\ \bibnamefont {{Sheikh-Jabbari}}},\ }\href {\doibase 10.1093/mnrasl/slad165} {\bibfield  {journal} {\bibinfo  {journal} {\mnras}\ }\textbf {\bibinfo {volume} {528}},\ \bibinfo {pages} {L20} (\bibinfo {year} {2024})},\ \Eprint {http://arxiv.org/abs/2303.06928} {arXiv:2303.06928 [astro-ph.CO]} \BibitemShut {NoStop}%
\bibitem [{\citenamefont {{Colg{\'a}in}}\ \emph {et~al.}(2022)\citenamefont {{Colg{\'a}in}}, \citenamefont {{Sheikh-Jabbari}},\ and\ \citenamefont {{Solomon}}}]{2022arXiv221102129C}%
  \BibitemOpen
  \bibfield  {author} {\bibinfo {author} {\bibfnamefont {E.~{\'O}.}\ \bibnamefont {{Colg{\'a}in}}}, \bibinfo {author} {\bibfnamefont {M.~M.}\ \bibnamefont {{Sheikh-Jabbari}}}, \ and\ \bibinfo {author} {\bibfnamefont {R.}~\bibnamefont {{Solomon}}},\ }\href {\doibase 10.48550/arXiv.2211.02129} {\bibfield  {journal} {\bibinfo  {journal} {arXiv e-prints}\ ,\ \bibinfo {eid} {arXiv:2211.02129}} (\bibinfo {year} {2022})},\ \Eprint {http://arxiv.org/abs/2211.02129} {arXiv:2211.02129 [astro-ph.CO]} \BibitemShut {NoStop}%
\bibitem [{\citenamefont {{{\'O} Colg{\'a}in}}\ \emph {et~al.}(2024)\citenamefont {{{\'O} Colg{\'a}in}}, \citenamefont {{Sheikh-Jabbari}}, \citenamefont {{Solomon}}, \citenamefont {{Dainotti}},\ and\ \citenamefont {{Stojkovic}}}]{2024PDU....4401464O}%
  \BibitemOpen
  \bibfield  {author} {\bibinfo {author} {\bibfnamefont {E.}~\bibnamefont {{{\'O} Colg{\'a}in}}}, \bibinfo {author} {\bibfnamefont {M.~M.}\ \bibnamefont {{Sheikh-Jabbari}}}, \bibinfo {author} {\bibfnamefont {R.}~\bibnamefont {{Solomon}}}, \bibinfo {author} {\bibfnamefont {M.~G.}\ \bibnamefont {{Dainotti}}}, \ and\ \bibinfo {author} {\bibfnamefont {D.}~\bibnamefont {{Stojkovic}}},\ }\href {\doibase 10.1016/j.dark.2024.101464} {\bibfield  {journal} {\bibinfo  {journal} {Physics of the Dark Universe}\ }\textbf {\bibinfo {volume} {44}},\ \bibinfo {eid} {101464} (\bibinfo {year} {2024})},\ \Eprint {http://arxiv.org/abs/2206.11447} {arXiv:2206.11447 [astro-ph.CO]} \BibitemShut {NoStop}%
\bibitem [{\citenamefont {{{\'O} Colg{\'a}in}}\ \emph {et~al.}(2022)\citenamefont {{{\'O} Colg{\'a}in}}, \citenamefont {{Sheikh-Jabbari}}, \citenamefont {{Solomon}}, \citenamefont {{Bargiacchi}}, \citenamefont {{Capozziello}}, \citenamefont {{Dainotti}},\ and\ \citenamefont {{Stojkovic}}}]{2022PhRvD.106d1301O}%
  \BibitemOpen
  \bibfield  {author} {\bibinfo {author} {\bibfnamefont {E.}~\bibnamefont {{{\'O} Colg{\'a}in}}}, \bibinfo {author} {\bibfnamefont {M.~M.}\ \bibnamefont {{Sheikh-Jabbari}}}, \bibinfo {author} {\bibfnamefont {R.}~\bibnamefont {{Solomon}}}, \bibinfo {author} {\bibfnamefont {G.}~\bibnamefont {{Bargiacchi}}}, \bibinfo {author} {\bibfnamefont {S.}~\bibnamefont {{Capozziello}}}, \bibinfo {author} {\bibfnamefont {M.~G.}\ \bibnamefont {{Dainotti}}}, \ and\ \bibinfo {author} {\bibfnamefont {D.}~\bibnamefont {{Stojkovic}}},\ }\href {\doibase 10.1103/PhysRevD.106.L041301} {\bibfield  {journal} {\bibinfo  {journal} {\prd}\ }\textbf {\bibinfo {volume} {106}},\ \bibinfo {eid} {L041301} (\bibinfo {year} {2022})},\ \Eprint {http://arxiv.org/abs/2203.10558} {arXiv:2203.10558 [astro-ph.CO]} \BibitemShut {NoStop}%
\bibitem [{\citenamefont {{Luongo}}\ \emph {et~al.}(2015)\citenamefont {{Luongo}}, \citenamefont {{Battista Pisani}},\ and\ \citenamefont {{Troisi}}}]{2015arXiv151207076L}%
  \BibitemOpen
  \bibfield  {author} {\bibinfo {author} {\bibfnamefont {O.}~\bibnamefont {{Luongo}}}, \bibinfo {author} {\bibfnamefont {G.}~\bibnamefont {{Battista Pisani}}}, \ and\ \bibinfo {author} {\bibfnamefont {A.}~\bibnamefont {{Troisi}}},\ }\href {\doibase 10.48550/arXiv.1512.07076} {\bibfield  {journal} {\bibinfo  {journal} {arXiv e-prints}\ ,\ \bibinfo {eid} {arXiv:1512.07076}} (\bibinfo {year} {2015})},\ \Eprint {http://arxiv.org/abs/1512.07076} {arXiv:1512.07076 [gr-qc]} \BibitemShut {NoStop}%
\bibitem [{\citenamefont {Refsdal}(1964)}]{Refsdal:1964nw}%
  \BibitemOpen
  \bibfield  {author} {\bibinfo {author} {\bibfnamefont {S.}~\bibnamefont {Refsdal}},\ }\href@noop {} {\bibfield  {journal} {\bibinfo  {journal} {Mon. Not. Roy. Astron. Soc.}\ }\textbf {\bibinfo {volume} {128}},\ \bibinfo {pages} {307} (\bibinfo {year} {1964})}\BibitemShut {NoStop}%
\bibitem [{\citenamefont {Kelly}\ \emph {et~al.}(2015)\citenamefont {Kelly} \emph {et~al.}}]{Kelly:2014mwa}%
  \BibitemOpen
  \bibfield  {author} {\bibinfo {author} {\bibfnamefont {P.~L.}\ \bibnamefont {Kelly}} \emph {et~al.},\ }\href {\doibase 10.1126/science.aaa3350} {\bibfield  {journal} {\bibinfo  {journal} {Science}\ }\textbf {\bibinfo {volume} {347}},\ \bibinfo {pages} {1123} (\bibinfo {year} {2015})},\ \Eprint {http://arxiv.org/abs/1411.6009} {arXiv:1411.6009 [astro-ph.CO]} \BibitemShut {NoStop}%
\bibitem [{\citenamefont {Goobar}\ \emph {et~al.}(2017)\citenamefont {Goobar} \emph {et~al.}}]{Goobar:2016uuf}%
  \BibitemOpen
  \bibfield  {author} {\bibinfo {author} {\bibfnamefont {A.}~\bibnamefont {Goobar}} \emph {et~al.},\ }\href {\doibase 10.1126/science.aal2729} {\bibfield  {journal} {\bibinfo  {journal} {Science}\ }\textbf {\bibinfo {volume} {356}},\ \bibinfo {pages} {291} (\bibinfo {year} {2017})},\ \Eprint {http://arxiv.org/abs/1611.00014} {arXiv:1611.00014 [astro-ph.CO]} \BibitemShut {NoStop}%
\bibitem [{\citenamefont {{Wolf}}\ and\ \citenamefont {{Ferreira}}(2023)}]{2023PhRvD.108j3519W}%
  \BibitemOpen
  \bibfield  {author} {\bibinfo {author} {\bibfnamefont {W.~J.}\ \bibnamefont {{Wolf}}}\ and\ \bibinfo {author} {\bibfnamefont {P.~G.}\ \bibnamefont {{Ferreira}}},\ }\href {\doibase 10.1103/PhysRevD.108.103519} {\bibfield  {journal} {\bibinfo  {journal} {\prd}\ }\textbf {\bibinfo {volume} {108}},\ \bibinfo {eid} {103519} (\bibinfo {year} {2023})},\ \Eprint {http://arxiv.org/abs/2310.07482} {arXiv:2310.07482 [astro-ph.CO]} \BibitemShut {NoStop}%
\bibitem [{\citenamefont {{Capozziello}}\ \emph {et~al.}(2018)\citenamefont {{Capozziello}}, \citenamefont {{Luongo}}, \citenamefont {{Pincak}},\ and\ \citenamefont {{Ravanpak}}}]{2018GReGr..50...53C}%
  \BibitemOpen
  \bibfield  {author} {\bibinfo {author} {\bibfnamefont {S.}~\bibnamefont {{Capozziello}}}, \bibinfo {author} {\bibfnamefont {O.}~\bibnamefont {{Luongo}}}, \bibinfo {author} {\bibfnamefont {R.}~\bibnamefont {{Pincak}}}, \ and\ \bibinfo {author} {\bibfnamefont {A.}~\bibnamefont {{Ravanpak}}},\ }\href {\doibase 10.1007/s10714-018-2374-4} {\bibfield  {journal} {\bibinfo  {journal} {General Relativity and Gravitation}\ }\textbf {\bibinfo {volume} {50}},\ \bibinfo {eid} {53} (\bibinfo {year} {2018})},\ \Eprint {http://arxiv.org/abs/1804.03649} {arXiv:1804.03649 [gr-qc]} \BibitemShut {NoStop}%
\bibitem [{\citenamefont {Capozziello}\ \emph {et~al.}(2019)\citenamefont {Capozziello}, \citenamefont {D'Agostino},\ and\ \citenamefont {Luongo}}]{Capozziello:2019cav}%
  \BibitemOpen
  \bibfield  {author} {\bibinfo {author} {\bibfnamefont {S.}~\bibnamefont {Capozziello}}, \bibinfo {author} {\bibfnamefont {R.}~\bibnamefont {D'Agostino}}, \ and\ \bibinfo {author} {\bibfnamefont {O.}~\bibnamefont {Luongo}},\ }\href {\doibase 10.1142/S0218271819300167} {\bibfield  {journal} {\bibinfo  {journal} {Int. J. Mod. Phys. D}\ }\textbf {\bibinfo {volume} {28}},\ \bibinfo {pages} {1930016} (\bibinfo {year} {2019})},\ \Eprint {http://arxiv.org/abs/1904.01427} {arXiv:1904.01427 [gr-qc]} \BibitemShut {NoStop}%
\bibitem [{\citenamefont {Rasanen}\ \emph {et~al.}(2015)\citenamefont {Rasanen}, \citenamefont {Bolejko},\ and\ \citenamefont {Finoguenov}}]{Rasanen:2014mca}%
  \BibitemOpen
  \bibfield  {author} {\bibinfo {author} {\bibfnamefont {S.}~\bibnamefont {Rasanen}}, \bibinfo {author} {\bibfnamefont {K.}~\bibnamefont {Bolejko}}, \ and\ \bibinfo {author} {\bibfnamefont {A.}~\bibnamefont {Finoguenov}},\ }\href {\doibase 10.1103/PhysRevLett.115.101301} {\bibfield  {journal} {\bibinfo  {journal} {Phys. Rev. Lett.}\ }\textbf {\bibinfo {volume} {115}},\ \bibinfo {pages} {101301} (\bibinfo {year} {2015})},\ \Eprint {http://arxiv.org/abs/1412.4976} {arXiv:1412.4976 [astro-ph.CO]} \BibitemShut {NoStop}%
\bibitem [{\citenamefont {Qi}\ \emph {et~al.}(2019{\natexlab{a}})\citenamefont {Qi}, \citenamefont {Cao}, \citenamefont {Biesiada}, \citenamefont {Ding}, \citenamefont {Zhu},\ and\ \citenamefont {Zheng}}]{Qi:2018atg}%
  \BibitemOpen
  \bibfield  {author} {\bibinfo {author} {\bibfnamefont {J.}~\bibnamefont {Qi}}, \bibinfo {author} {\bibfnamefont {S.}~\bibnamefont {Cao}}, \bibinfo {author} {\bibfnamefont {M.}~\bibnamefont {Biesiada}}, \bibinfo {author} {\bibfnamefont {X.}~\bibnamefont {Ding}}, \bibinfo {author} {\bibfnamefont {Z.-H.}\ \bibnamefont {Zhu}}, \ and\ \bibinfo {author} {\bibfnamefont {X.}~\bibnamefont {Zheng}},\ }\href {\doibase 10.1103/PhysRevD.100.023530} {\bibfield  {journal} {\bibinfo  {journal} {Phys. Rev. D}\ }\textbf {\bibinfo {volume} {100}},\ \bibinfo {pages} {023530} (\bibinfo {year} {2019}{\natexlab{a}})},\ \Eprint {http://arxiv.org/abs/1802.05532} {arXiv:1802.05532 [astro-ph.CO]} \BibitemShut {NoStop}%
\bibitem [{\citenamefont {Wang}\ \emph {et~al.}(2020)\citenamefont {Wang}, \citenamefont {Qi}, \citenamefont {Zhang},\ and\ \citenamefont {Zhang}}]{Wang:2019yob}%
  \BibitemOpen
  \bibfield  {author} {\bibinfo {author} {\bibfnamefont {B.}~\bibnamefont {Wang}}, \bibinfo {author} {\bibfnamefont {J.-Z.}\ \bibnamefont {Qi}}, \bibinfo {author} {\bibfnamefont {J.-F.}\ \bibnamefont {Zhang}}, \ and\ \bibinfo {author} {\bibfnamefont {X.}~\bibnamefont {Zhang}},\ }\href@noop {} {\bibfield  {journal} {\bibinfo  {journal} {The Astrophysical Journal}\ }\textbf {\bibinfo {volume} {898}},\ \bibinfo {pages} {100} (\bibinfo {year} {2020})}\BibitemShut {NoStop}%
\bibitem [{\citenamefont {Xia}\ \emph {et~al.}(2017)\citenamefont {Xia}, \citenamefont {Yu}, \citenamefont {Wang}, \citenamefont {Tian}, \citenamefont {Li}, \citenamefont {Cao},\ and\ \citenamefont {Zhu}}]{Xia:2016dgk}%
  \BibitemOpen
  \bibfield  {author} {\bibinfo {author} {\bibfnamefont {J.-Q.}\ \bibnamefont {Xia}}, \bibinfo {author} {\bibfnamefont {H.}~\bibnamefont {Yu}}, \bibinfo {author} {\bibfnamefont {G.-J.}\ \bibnamefont {Wang}}, \bibinfo {author} {\bibfnamefont {S.-X.}\ \bibnamefont {Tian}}, \bibinfo {author} {\bibfnamefont {Z.-X.}\ \bibnamefont {Li}}, \bibinfo {author} {\bibfnamefont {S.}~\bibnamefont {Cao}}, \ and\ \bibinfo {author} {\bibfnamefont {Z.-H.}\ \bibnamefont {Zhu}},\ }\href {\doibase 10.3847/1538-4357/834/1/75} {\bibfield  {journal} {\bibinfo  {journal} {Astrophys. J.}\ }\textbf {\bibinfo {volume} {834}},\ \bibinfo {pages} {75} (\bibinfo {year} {2017})},\ \Eprint {http://arxiv.org/abs/1611.04731} {arXiv:1611.04731 [astro-ph.CO]} \BibitemShut {NoStop}%
\bibitem [{\citenamefont {Li}\ \emph {et~al.}(2018)\citenamefont {Li}, \citenamefont {Ding}, \citenamefont {Wang}, \citenamefont {Liao},\ and\ \citenamefont {Zhu}}]{Li:2018hyr}%
  \BibitemOpen
  \bibfield  {author} {\bibinfo {author} {\bibfnamefont {Z.}~\bibnamefont {Li}}, \bibinfo {author} {\bibfnamefont {X.}~\bibnamefont {Ding}}, \bibinfo {author} {\bibfnamefont {G.-J.}\ \bibnamefont {Wang}}, \bibinfo {author} {\bibfnamefont {K.}~\bibnamefont {Liao}}, \ and\ \bibinfo {author} {\bibfnamefont {Z.-H.}\ \bibnamefont {Zhu}},\ }\href {\doibase 10.3847/1538-4357/aaa76f} {\bibfield  {journal} {\bibinfo  {journal} {Astrophys. J.}\ }\textbf {\bibinfo {volume} {854}},\ \bibinfo {pages} {146} (\bibinfo {year} {2018})},\ \Eprint {http://arxiv.org/abs/1801.08001} {arXiv:1801.08001 [astro-ph.CO]} \BibitemShut {NoStop}%
\bibitem [{\citenamefont {Zhou}\ and\ \citenamefont {Li}(2020)}]{Zhou:2019vou}%
  \BibitemOpen
  \bibfield  {author} {\bibinfo {author} {\bibfnamefont {H.}~\bibnamefont {Zhou}}\ and\ \bibinfo {author} {\bibfnamefont {Z.-X.}\ \bibnamefont {Li}},\ }\href {\doibase 10.3847/1538-4357/ab5f61} {\bibfield  {journal} {\bibinfo  {journal} {Astrophys. J.}\ }\textbf {\bibinfo {volume} {899}},\ \bibinfo {pages} {186} (\bibinfo {year} {2020})},\ \Eprint {http://arxiv.org/abs/1912.01828} {arXiv:1912.01828 [astro-ph.CO]} \BibitemShut {NoStop}%
\bibitem [{\citenamefont {Cao}\ \emph {et~al.}(2015)\citenamefont {Cao}, \citenamefont {Biesiada}, \citenamefont {Gavazzi}, \citenamefont {Pi{\'o}rkowska},\ and\ \citenamefont {Zhu}}]{Cao:2015qja}%
  \BibitemOpen
  \bibfield  {author} {\bibinfo {author} {\bibfnamefont {S.}~\bibnamefont {Cao}}, \bibinfo {author} {\bibfnamefont {M.}~\bibnamefont {Biesiada}}, \bibinfo {author} {\bibfnamefont {R.}~\bibnamefont {Gavazzi}}, \bibinfo {author} {\bibfnamefont {A.}~\bibnamefont {Pi{\'o}rkowska}}, \ and\ \bibinfo {author} {\bibfnamefont {Z.-H.}\ \bibnamefont {Zhu}},\ }\href@noop {} {\bibfield  {journal} {\bibinfo  {journal} {The Astrophysical Journal}\ }\textbf {\bibinfo {volume} {806}},\ \bibinfo {pages} {185} (\bibinfo {year} {2015})}\BibitemShut {NoStop}%
\bibitem [{\citenamefont {Chen}\ \emph {et~al.}(2019)\citenamefont {Chen}, \citenamefont {Li}, \citenamefont {Shu},\ and\ \citenamefont {Cao}}]{Chen:2018jcf}%
  \BibitemOpen
  \bibfield  {author} {\bibinfo {author} {\bibfnamefont {Y.}~\bibnamefont {Chen}}, \bibinfo {author} {\bibfnamefont {R.}~\bibnamefont {Li}}, \bibinfo {author} {\bibfnamefont {Y.}~\bibnamefont {Shu}}, \ and\ \bibinfo {author} {\bibfnamefont {X.}~\bibnamefont {Cao}},\ }\href {\doibase 10.1093/mnras/stz1902} {\bibfield  {journal} {\bibinfo  {journal} {Mon. Not. Roy. Astron. Soc.}\ }\textbf {\bibinfo {volume} {488}},\ \bibinfo {pages} {3745} (\bibinfo {year} {2019})},\ \Eprint {http://arxiv.org/abs/1809.09845} {arXiv:1809.09845 [astro-ph.CO]} \BibitemShut {NoStop}%
\bibitem [{\citenamefont {Qi}\ \emph {et~al.}(2019{\natexlab{b}})\citenamefont {Qi}, \citenamefont {Cao}, \citenamefont {Zhang}, \citenamefont {Biesiada}, \citenamefont {Wu},\ and\ \citenamefont {Zhu}}]{Qi:2018aio}%
  \BibitemOpen
  \bibfield  {author} {\bibinfo {author} {\bibfnamefont {J.-Z.}\ \bibnamefont {Qi}}, \bibinfo {author} {\bibfnamefont {S.}~\bibnamefont {Cao}}, \bibinfo {author} {\bibfnamefont {S.}~\bibnamefont {Zhang}}, \bibinfo {author} {\bibfnamefont {M.}~\bibnamefont {Biesiada}}, \bibinfo {author} {\bibfnamefont {Y.}~\bibnamefont {Wu}}, \ and\ \bibinfo {author} {\bibfnamefont {Z.-H.}\ \bibnamefont {Zhu}},\ }\href {\doibase 10.1093/mnras/sty3175} {\bibfield  {journal} {\bibinfo  {journal} {Mon. Not. Roy. Astron. Soc.}\ }\textbf {\bibinfo {volume} {483}},\ \bibinfo {pages} {1104} (\bibinfo {year} {2019}{\natexlab{b}})},\ \Eprint {http://arxiv.org/abs/1803.01990} {arXiv:1803.01990 [astro-ph.CO]} \BibitemShut {NoStop}%
\bibitem [{\citenamefont {Liao}(2019)}]{Liao:2019hfl}%
  \BibitemOpen
  \bibfield  {author} {\bibinfo {author} {\bibfnamefont {K.}~\bibnamefont {Liao}},\ }\href {\doibase 10.1103/PhysRevD.99.083514} {\bibfield  {journal} {\bibinfo  {journal} {Phys. Rev. D}\ }\textbf {\bibinfo {volume} {99}},\ \bibinfo {pages} {083514} (\bibinfo {year} {2019})},\ \Eprint {http://arxiv.org/abs/1904.01744} {arXiv:1904.01744 [astro-ph.CO]} \BibitemShut {NoStop}%
\bibitem [{\citenamefont {Cao}\ \emph {et~al.}(2012)\citenamefont {Cao}, \citenamefont {Pan}, \citenamefont {Biesiada}, \citenamefont {Godlowski},\ and\ \citenamefont {Zhu}}]{cao2012constraints}%
  \BibitemOpen
  \bibfield  {author} {\bibinfo {author} {\bibfnamefont {S.}~\bibnamefont {Cao}}, \bibinfo {author} {\bibfnamefont {Y.}~\bibnamefont {Pan}}, \bibinfo {author} {\bibfnamefont {M.}~\bibnamefont {Biesiada}}, \bibinfo {author} {\bibfnamefont {W.}~\bibnamefont {Godlowski}}, \ and\ \bibinfo {author} {\bibfnamefont {Z.}~\bibnamefont {Zhu}},\ }\href@noop {} {\bibfield  {journal} {\bibinfo  {journal} {Journal of Cosmology and Astroparticle Physics}\ }\textbf {\bibinfo {volume} {2012}},\ \bibinfo {pages} {016} (\bibinfo {year} {2012})}\BibitemShut {NoStop}%
\bibitem [{\citenamefont {{Amati}}\ \emph {et~al.}(2019)\citenamefont {{Amati}}, \citenamefont {{D'Agostino}}, \citenamefont {{Luongo}}, \citenamefont {{Muccino}},\ and\ \citenamefont {{Tantalo}}}]{2019MNRAS.486L..46A}%
  \BibitemOpen
  \bibfield  {author} {\bibinfo {author} {\bibfnamefont {L.}~\bibnamefont {{Amati}}}, \bibinfo {author} {\bibfnamefont {R.}~\bibnamefont {{D'Agostino}}}, \bibinfo {author} {\bibfnamefont {O.}~\bibnamefont {{Luongo}}}, \bibinfo {author} {\bibfnamefont {M.}~\bibnamefont {{Muccino}}}, \ and\ \bibinfo {author} {\bibfnamefont {M.}~\bibnamefont {{Tantalo}}},\ }\href {\doibase 10.1093/mnrasl/slz056} {\bibfield  {journal} {\bibinfo  {journal} {\mnras}\ }\textbf {\bibinfo {volume} {486}},\ \bibinfo {pages} {L46} (\bibinfo {year} {2019})},\ \Eprint {http://arxiv.org/abs/1811.08934} {arXiv:1811.08934 [astro-ph.HE]} \BibitemShut {NoStop}%
\bibitem [{\citenamefont {{Arjona}}\ \emph {et~al.}(2019)\citenamefont {{Arjona}}, \citenamefont {{Cardona}},\ and\ \citenamefont {{Nesseris}}}]{2019PhRvD..99d3516A}%
  \BibitemOpen
  \bibfield  {author} {\bibinfo {author} {\bibfnamefont {R.}~\bibnamefont {{Arjona}}}, \bibinfo {author} {\bibfnamefont {W.}~\bibnamefont {{Cardona}}}, \ and\ \bibinfo {author} {\bibfnamefont {S.}~\bibnamefont {{Nesseris}}},\ }\href {\doibase 10.1103/PhysRevD.99.043516} {\bibfield  {journal} {\bibinfo  {journal} {\prd}\ }\textbf {\bibinfo {volume} {99}},\ \bibinfo {eid} {043516} (\bibinfo {year} {2019})},\ \Eprint {http://arxiv.org/abs/1811.02469} {arXiv:1811.02469 [astro-ph.CO]} \BibitemShut {NoStop}%
\bibitem [{\citenamefont {{Luongo}}\ \emph {et~al.}(2022)\citenamefont {{Luongo}}, \citenamefont {{Muccino}}, \citenamefont {{Colg{\'a}in}}, \citenamefont {{Sheikh-Jabbari}},\ and\ \citenamefont {{Yin}}}]{2022PhRvD.105j3510L}%
  \BibitemOpen
  \bibfield  {author} {\bibinfo {author} {\bibfnamefont {O.}~\bibnamefont {{Luongo}}}, \bibinfo {author} {\bibfnamefont {M.}~\bibnamefont {{Muccino}}}, \bibinfo {author} {\bibfnamefont {E.~{\'O}.}\ \bibnamefont {{Colg{\'a}in}}}, \bibinfo {author} {\bibfnamefont {M.~M.}\ \bibnamefont {{Sheikh-Jabbari}}}, \ and\ \bibinfo {author} {\bibfnamefont {L.}~\bibnamefont {{Yin}}},\ }\href {\doibase 10.1103/PhysRevD.105.103510} {\bibfield  {journal} {\bibinfo  {journal} {\prd}\ }\textbf {\bibinfo {volume} {105}},\ \bibinfo {eid} {103510} (\bibinfo {year} {2022})},\ \Eprint {http://arxiv.org/abs/2108.13228} {arXiv:2108.13228 [astro-ph.CO]} \BibitemShut {NoStop}%
\bibitem [{\citenamefont {Cao}\ \emph {et~al.}(2019)\citenamefont {Cao}, \citenamefont {Qi}, \citenamefont {Cao}, \citenamefont {Biesiada}, \citenamefont {Li}, \citenamefont {Pan},\ and\ \citenamefont {Zhu}}]{cao2019direct}%
  \BibitemOpen
  \bibfield  {author} {\bibinfo {author} {\bibfnamefont {S.}~\bibnamefont {Cao}}, \bibinfo {author} {\bibfnamefont {J.}~\bibnamefont {Qi}}, \bibinfo {author} {\bibfnamefont {Z.}~\bibnamefont {Cao}}, \bibinfo {author} {\bibfnamefont {M.}~\bibnamefont {Biesiada}}, \bibinfo {author} {\bibfnamefont {J.}~\bibnamefont {Li}}, \bibinfo {author} {\bibfnamefont {Y.}~\bibnamefont {Pan}}, \ and\ \bibinfo {author} {\bibfnamefont {Z.}~\bibnamefont {Zhu}},\ }\href@noop {} {\bibfield  {journal} {\bibinfo  {journal} {Scientific Reports}\ }\textbf {\bibinfo {volume} {9}},\ \bibinfo {pages} {11608} (\bibinfo {year} {2019})}\BibitemShut {NoStop}%
\bibitem [{\citenamefont {{Luongo}}\ and\ \citenamefont {{Muccino}}(2024{\natexlab{b}})}]{2024arXiv241104901L}%
  \BibitemOpen
  \bibfield  {author} {\bibinfo {author} {\bibfnamefont {O.}~\bibnamefont {{Luongo}}}\ and\ \bibinfo {author} {\bibfnamefont {M.}~\bibnamefont {{Muccino}}},\ }\href {\doibase 10.48550/arXiv.2411.04901} {\bibfield  {journal} {\bibinfo  {journal} {arXiv e-prints}\ ,\ \bibinfo {eid} {arXiv:2411.04901}} (\bibinfo {year} {2024}{\natexlab{b}})},\ \Eprint {http://arxiv.org/abs/2411.04901} {arXiv:2411.04901 [astro-ph.CO]} \BibitemShut {NoStop}%
\bibitem [{\citenamefont {{Jimenez}}\ and\ \citenamefont {{Loeb}}(2002)}]{2002ApJ...573...37J}%
  \BibitemOpen
  \bibfield  {author} {\bibinfo {author} {\bibfnamefont {R.}~\bibnamefont {{Jimenez}}}\ and\ \bibinfo {author} {\bibfnamefont {A.}~\bibnamefont {{Loeb}}},\ }\href {\doibase 10.1086/340549} {\bibfield  {journal} {\bibinfo  {journal} {\apj}\ }\textbf {\bibinfo {volume} {573}},\ \bibinfo {pages} {37} (\bibinfo {year} {2002})},\ \Eprint {http://arxiv.org/abs/astro-ph/0106145} {astro-ph/0106145} \BibitemShut {NoStop}%
\bibitem [{\citenamefont {{Moresco}}\ \emph {et~al.}(2022)\citenamefont {{Moresco}}, \citenamefont {{Amati}}, \citenamefont {{Amendola}}, \citenamefont {{Birrer}}, \citenamefont {{Blakeslee}}, \citenamefont {{Cantiello}}, \citenamefont {{Cimatti}}, \citenamefont {{Darling}}, \citenamefont {{Della Valle}}, \citenamefont {{Fishbach}}, \citenamefont {{Grillo}}, \citenamefont {{Hamaus}}, \citenamefont {{Holz}}, \citenamefont {{Izzo}}, \citenamefont {{Jimenez}}, \citenamefont {{Lusso}}, \citenamefont {{Meneghetti}}, \citenamefont {{Piedipalumbo}}, \citenamefont {{Pisani}}, \citenamefont {{Pourtsidou}}, \citenamefont {{Pozzetti}}, \citenamefont {{Quartin}}, \citenamefont {{Risaliti}}, \citenamefont {{Rosati}},\ and\ \citenamefont {{Verde}}}]{2022LRR....25....6M}%
  \BibitemOpen
  \bibfield  {author} {\bibinfo {author} {\bibfnamefont {M.}~\bibnamefont {{Moresco}}}, \bibinfo {author} {\bibfnamefont {L.}~\bibnamefont {{Amati}}}, \bibinfo {author} {\bibfnamefont {L.}~\bibnamefont {{Amendola}}}, \bibinfo {author} {\bibfnamefont {S.}~\bibnamefont {{Birrer}}}, \bibinfo {author} {\bibfnamefont {J.~P.}\ \bibnamefont {{Blakeslee}}}, \bibinfo {author} {\bibfnamefont {M.}~\bibnamefont {{Cantiello}}}, \bibinfo {author} {\bibfnamefont {A.}~\bibnamefont {{Cimatti}}}, \bibinfo {author} {\bibfnamefont {J.}~\bibnamefont {{Darling}}}, \bibinfo {author} {\bibfnamefont {M.}~\bibnamefont {{Della Valle}}}, \bibinfo {author} {\bibfnamefont {M.}~\bibnamefont {{Fishbach}}}, \bibinfo {author} {\bibfnamefont {C.}~\bibnamefont {{Grillo}}}, \bibinfo {author} {\bibfnamefont {N.}~\bibnamefont {{Hamaus}}}, \bibinfo {author} {\bibfnamefont {D.}~\bibnamefont {{Holz}}}, \bibinfo {author} {\bibfnamefont {L.}~\bibnamefont {{Izzo}}}, \bibinfo {author} {\bibfnamefont {R.}~\bibnamefont {{Jimenez}}}, \bibinfo {author}
  {\bibfnamefont {E.}~\bibnamefont {{Lusso}}}, \bibinfo {author} {\bibfnamefont {M.}~\bibnamefont {{Meneghetti}}}, \bibinfo {author} {\bibfnamefont {E.}~\bibnamefont {{Piedipalumbo}}}, \bibinfo {author} {\bibfnamefont {A.}~\bibnamefont {{Pisani}}}, \bibinfo {author} {\bibfnamefont {A.}~\bibnamefont {{Pourtsidou}}}, \bibinfo {author} {\bibfnamefont {L.}~\bibnamefont {{Pozzetti}}}, \bibinfo {author} {\bibfnamefont {M.}~\bibnamefont {{Quartin}}}, \bibinfo {author} {\bibfnamefont {G.}~\bibnamefont {{Risaliti}}}, \bibinfo {author} {\bibfnamefont {P.}~\bibnamefont {{Rosati}}}, \ and\ \bibinfo {author} {\bibfnamefont {L.}~\bibnamefont {{Verde}}},\ }\href {\doibase 10.1007/s41114-022-00040-z} {\bibfield  {journal} {\bibinfo  {journal} {Living Reviews in Relativity}\ }\textbf {\bibinfo {volume} {25}},\ \bibinfo {eid} {6} (\bibinfo {year} {2022})},\ \Eprint {http://arxiv.org/abs/2201.07241} {arXiv:2201.07241 [astro-ph.CO]} \BibitemShut {NoStop}%
\bibitem [{\citenamefont {{Luongo}}\ and\ \citenamefont {{Muccino}}(2021{\natexlab{a}})}]{2021MNRAS.503.4581L}%
  \BibitemOpen
  \bibfield  {author} {\bibinfo {author} {\bibfnamefont {O.}~\bibnamefont {{Luongo}}}\ and\ \bibinfo {author} {\bibfnamefont {M.}~\bibnamefont {{Muccino}}},\ }\href {\doibase 10.1093/mnras/stab795} {\bibfield  {journal} {\bibinfo  {journal} {\mnras}\ }\textbf {\bibinfo {volume} {503}},\ \bibinfo {pages} {4581} (\bibinfo {year} {2021}{\natexlab{a}})},\ \Eprint {http://arxiv.org/abs/2011.13590} {arXiv:2011.13590 [astro-ph.CO]} \BibitemShut {NoStop}%
\bibitem [{\citenamefont {{Montiel}}\ \emph {et~al.}(2021)\citenamefont {{Montiel}}, \citenamefont {{Cabrera}},\ and\ \citenamefont {{Hidalgo}}}]{2021MNRAS.501.3515M}%
  \BibitemOpen
  \bibfield  {author} {\bibinfo {author} {\bibfnamefont {A.}~\bibnamefont {{Montiel}}}, \bibinfo {author} {\bibfnamefont {J.~I.}\ \bibnamefont {{Cabrera}}}, \ and\ \bibinfo {author} {\bibfnamefont {J.~C.}\ \bibnamefont {{Hidalgo}}},\ }\href {\doibase 10.1093/mnras/staa3926} {\bibfield  {journal} {\bibinfo  {journal} {\mnras}\ }\textbf {\bibinfo {volume} {501}},\ \bibinfo {pages} {3515} (\bibinfo {year} {2021})},\ \Eprint {http://arxiv.org/abs/2003.03387} {arXiv:2003.03387 [astro-ph.HE]} \BibitemShut {NoStop}%
\bibitem [{\citenamefont {{Luongo}}\ and\ \citenamefont {{Muccino}}(2023)}]{2023MNRAS.518.2247L}%
  \BibitemOpen
  \bibfield  {author} {\bibinfo {author} {\bibfnamefont {O.}~\bibnamefont {{Luongo}}}\ and\ \bibinfo {author} {\bibfnamefont {M.}~\bibnamefont {{Muccino}}},\ }\href {\doibase 10.1093/mnras/stac2925} {\bibfield  {journal} {\bibinfo  {journal} {\mnras}\ }\textbf {\bibinfo {volume} {518}},\ \bibinfo {pages} {2247} (\bibinfo {year} {2023})},\ \Eprint {http://arxiv.org/abs/2207.00440} {arXiv:2207.00440 [astro-ph.CO]} \BibitemShut {NoStop}%
\bibitem [{\citenamefont {{Muccino}}\ \emph {et~al.}(2023)\citenamefont {{Muccino}}, \citenamefont {{Luongo}},\ and\ \citenamefont {{Jain}}}]{2023MNRAS.523.4938M}%
  \BibitemOpen
  \bibfield  {author} {\bibinfo {author} {\bibfnamefont {M.}~\bibnamefont {{Muccino}}}, \bibinfo {author} {\bibfnamefont {O.}~\bibnamefont {{Luongo}}}, \ and\ \bibinfo {author} {\bibfnamefont {D.}~\bibnamefont {{Jain}}},\ }\href {\doibase 10.1093/mnras/stad1760} {\bibfield  {journal} {\bibinfo  {journal} {\mnras}\ }\textbf {\bibinfo {volume} {523}},\ \bibinfo {pages} {4938} (\bibinfo {year} {2023})},\ \Eprint {http://arxiv.org/abs/2208.13700} {arXiv:2208.13700 [astro-ph.CO]} \BibitemShut {NoStop}%
\bibitem [{\citenamefont {{Alfano}}\ \emph {et~al.}(2024{\natexlab{a}})\citenamefont {{Alfano}}, \citenamefont {{Capozziello}}, \citenamefont {{Luongo}},\ and\ \citenamefont {{Muccino}}}]{2024JHEAp..42..178A}%
  \BibitemOpen
  \bibfield  {author} {\bibinfo {author} {\bibfnamefont {A.~C.}\ \bibnamefont {{Alfano}}}, \bibinfo {author} {\bibfnamefont {S.}~\bibnamefont {{Capozziello}}}, \bibinfo {author} {\bibfnamefont {O.}~\bibnamefont {{Luongo}}}, \ and\ \bibinfo {author} {\bibfnamefont {M.}~\bibnamefont {{Muccino}}},\ }\href {\doibase 10.1016/j.jheap.2024.05.002} {\bibfield  {journal} {\bibinfo  {journal} {Journal of High Energy Astrophysics}\ }\textbf {\bibinfo {volume} {42}},\ \bibinfo {pages} {178} (\bibinfo {year} {2024}{\natexlab{a}})},\ \Eprint {http://arxiv.org/abs/2402.18967} {arXiv:2402.18967 [astro-ph.CO]} \BibitemShut {NoStop}%
\bibitem [{\citenamefont {{Alfano}}\ \emph {et~al.}(2024{\natexlab{b}})\citenamefont {{Alfano}}, \citenamefont {{Luongo}},\ and\ \citenamefont {{Muccino}}}]{2024A&A...686A..30A}%
  \BibitemOpen
  \bibfield  {author} {\bibinfo {author} {\bibfnamefont {A.~C.}\ \bibnamefont {{Alfano}}}, \bibinfo {author} {\bibfnamefont {O.}~\bibnamefont {{Luongo}}}, \ and\ \bibinfo {author} {\bibfnamefont {M.}~\bibnamefont {{Muccino}}},\ }\href {\doibase 10.1051/0004-6361/202348585} {\bibfield  {journal} {\bibinfo  {journal} {\aap}\ }\textbf {\bibinfo {volume} {686}},\ \bibinfo {eid} {A30} (\bibinfo {year} {2024}{\natexlab{b}})},\ \Eprint {http://arxiv.org/abs/2311.05324} {arXiv:2311.05324 [astro-ph.CO]} \BibitemShut {NoStop}%
\bibitem [{\citenamefont {{Alfano}}\ \emph {et~al.}(2024{\natexlab{c}})\citenamefont {{Alfano}}, \citenamefont {{Luongo}},\ and\ \citenamefont {{Muccino}}}]{2024arXiv240802536A}%
  \BibitemOpen
  \bibfield  {author} {\bibinfo {author} {\bibfnamefont {A.~C.}\ \bibnamefont {{Alfano}}}, \bibinfo {author} {\bibfnamefont {O.}~\bibnamefont {{Luongo}}}, \ and\ \bibinfo {author} {\bibfnamefont {M.}~\bibnamefont {{Muccino}}},\ }\href {\doibase 10.48550/arXiv.2408.02536} {\bibfield  {journal} {\bibinfo  {journal} {arXiv e-prints}\ ,\ \bibinfo {eid} {arXiv:2408.02536}} (\bibinfo {year} {2024}{\natexlab{c}})},\ \Eprint {http://arxiv.org/abs/2408.02536} {arXiv:2408.02536 [astro-ph.CO]} \BibitemShut {NoStop}%
\bibitem [{\citenamefont {{Amati}}\ and\ \citenamefont {{Della Valle}}(2013)}]{AmatiDellaValle2013}%
  \BibitemOpen
  \bibfield  {author} {\bibinfo {author} {\bibfnamefont {L.}~\bibnamefont {{Amati}}}\ and\ \bibinfo {author} {\bibfnamefont {M.}~\bibnamefont {{Della Valle}}},\ }\href {\doibase 10.1142/S0218271813300280} {\bibfield  {journal} {\bibinfo  {journal} {International Journal of Modern Physics D}\ }\textbf {\bibinfo {volume} {22}},\ \bibinfo {eid} {1330028} (\bibinfo {year} {2013})},\ \Eprint {http://arxiv.org/abs/1310.3141} {arXiv:1310.3141 [astro-ph.CO]} \BibitemShut {NoStop}%
\bibitem [{\citenamefont {{Yonetoku}}\ \emph {et~al.}(2004)\citenamefont {{Yonetoku}}, \citenamefont {{Murakami}}, \citenamefont {{Nakamura}}, \citenamefont {{Yamazaki}}, \citenamefont {{Inoue}},\ and\ \citenamefont {{Ioka}}}]{Yonetoku}%
  \BibitemOpen
  \bibfield  {author} {\bibinfo {author} {\bibfnamefont {D.}~\bibnamefont {{Yonetoku}}}, \bibinfo {author} {\bibfnamefont {T.}~\bibnamefont {{Murakami}}}, \bibinfo {author} {\bibfnamefont {T.}~\bibnamefont {{Nakamura}}}, \bibinfo {author} {\bibfnamefont {R.}~\bibnamefont {{Yamazaki}}}, \bibinfo {author} {\bibfnamefont {A.~K.}\ \bibnamefont {{Inoue}}}, \ and\ \bibinfo {author} {\bibfnamefont {K.}~\bibnamefont {{Ioka}}},\ }\href {\doibase 10.1086/421285} {\bibfield  {journal} {\bibinfo  {journal} {\apj}\ }\textbf {\bibinfo {volume} {609}},\ \bibinfo {pages} {935} (\bibinfo {year} {2004})},\ \Eprint {http://arxiv.org/abs/astro-ph/0309217} {arXiv:astro-ph/0309217 [astro-ph]} \BibitemShut {NoStop}%
\bibitem [{\citenamefont {{Ghirlanda}}\ \emph {et~al.}(2004)\citenamefont {{Ghirlanda}}, \citenamefont {{Ghisellini}},\ and\ \citenamefont {{Lazzati}}}]{ghirlanda}%
  \BibitemOpen
  \bibfield  {author} {\bibinfo {author} {\bibfnamefont {G.}~\bibnamefont {{Ghirlanda}}}, \bibinfo {author} {\bibfnamefont {G.}~\bibnamefont {{Ghisellini}}}, \ and\ \bibinfo {author} {\bibfnamefont {D.}~\bibnamefont {{Lazzati}}},\ }\href {\doibase 10.1086/424913} {\bibfield  {journal} {\bibinfo  {journal} {\apj}\ }\textbf {\bibinfo {volume} {616}},\ \bibinfo {pages} {331} (\bibinfo {year} {2004})},\ \Eprint {http://arxiv.org/abs/astro-ph/0405602} {arXiv:astro-ph/0405602 [astro-ph]} \BibitemShut {NoStop}%
\bibitem [{\citenamefont {{Muccino}}\ \emph {et~al.}(2021)\citenamefont {{Muccino}}, \citenamefont {{Izzo}}, \citenamefont {{Luongo}}, \citenamefont {{Boshkayev}}, \citenamefont {{Amati}} \emph {et~al.}}]{2021ApJ...908..181M}%
  \BibitemOpen
  \bibfield  {author} {\bibinfo {author} {\bibfnamefont {M.}~\bibnamefont {{Muccino}}}, \bibinfo {author} {\bibfnamefont {L.}~\bibnamefont {{Izzo}}}, \bibinfo {author} {\bibfnamefont {O.}~\bibnamefont {{Luongo}}}, \bibinfo {author} {\bibfnamefont {K.}~\bibnamefont {{Boshkayev}}}, \bibinfo {author} {\bibfnamefont {L.}~\bibnamefont {{Amati}}},  \emph {et~al.},\ }\href {\doibase 10.3847/1538-4357/abd254} {\bibfield  {journal} {\bibinfo  {journal} {\apj}\ }\textbf {\bibinfo {volume} {908}},\ \bibinfo {eid} {181} (\bibinfo {year} {2021})},\ \Eprint {http://arxiv.org/abs/2012.03392} {arXiv:2012.03392 [astro-ph.CO]} \BibitemShut {NoStop}%
\bibitem [{\citenamefont {{Luongo}}\ and\ \citenamefont {{Muccino}}(2021{\natexlab{b}})}]{2021Galax...9...77L}%
  \BibitemOpen
  \bibfield  {author} {\bibinfo {author} {\bibfnamefont {O.}~\bibnamefont {{Luongo}}}\ and\ \bibinfo {author} {\bibfnamefont {M.}~\bibnamefont {{Muccino}}},\ }\href {\doibase 10.3390/galaxies9040077} {\bibfield  {journal} {\bibinfo  {journal} {Galaxies}\ }\textbf {\bibinfo {volume} {9}},\ \bibinfo {pages} {77} (\bibinfo {year} {2021}{\natexlab{b}})},\ \Eprint {http://arxiv.org/abs/2110.14408} {arXiv:2110.14408 [astro-ph.HE]} \BibitemShut {NoStop}%
\bibitem [{\citenamefont {{Cao}}\ \emph {et~al.}(2022)\citenamefont {{Cao}}, \citenamefont {{Dainotti}},\ and\ \citenamefont {{Ratra}}}]{2022MNRAS.512..439C}%
  \BibitemOpen
  \bibfield  {author} {\bibinfo {author} {\bibfnamefont {S.}~\bibnamefont {{Cao}}}, \bibinfo {author} {\bibfnamefont {M.}~\bibnamefont {{Dainotti}}}, \ and\ \bibinfo {author} {\bibfnamefont {B.}~\bibnamefont {{Ratra}}},\ }\href {\doibase 10.1093/mnras/stac517} {\bibfield  {journal} {\bibinfo  {journal} {\mnras}\ }\textbf {\bibinfo {volume} {512}},\ \bibinfo {pages} {439} (\bibinfo {year} {2022})},\ \Eprint {http://arxiv.org/abs/2201.05245} {arXiv:2201.05245 [astro-ph.CO]} \BibitemShut {NoStop}%
\bibitem [{\citenamefont {{Khadka}}\ \emph {et~al.}(2021)\citenamefont {{Khadka}}, \citenamefont {{Luongo}}, \citenamefont {{Muccino}},\ and\ \citenamefont {{Ratra}}}]{2021JCAP...09..042K}%
  \BibitemOpen
  \bibfield  {author} {\bibinfo {author} {\bibfnamefont {N.}~\bibnamefont {{Khadka}}}, \bibinfo {author} {\bibfnamefont {O.}~\bibnamefont {{Luongo}}}, \bibinfo {author} {\bibfnamefont {M.}~\bibnamefont {{Muccino}}}, \ and\ \bibinfo {author} {\bibfnamefont {B.}~\bibnamefont {{Ratra}}},\ }\href {\doibase 10.1088/1475-7516/2021/09/042} {\bibfield  {journal} {\bibinfo  {journal} {\jcap}\ }\textbf {\bibinfo {volume} {2021}},\ \bibinfo {eid} {042} (\bibinfo {year} {2021})},\ \Eprint {http://arxiv.org/abs/2105.12692} {arXiv:2105.12692 [astro-ph.CO]} \BibitemShut {NoStop}%
\bibitem [{\citenamefont {{Qi}}\ \emph {et~al.}(2021)\citenamefont {{Qi}}, \citenamefont {{Zhao}}, \citenamefont {{Cao}}, \citenamefont {{Biesiada}},\ and\ \citenamefont {{Liu}}}]{2021MNRAS.503.2179Q}%
  \BibitemOpen
  \bibfield  {author} {\bibinfo {author} {\bibfnamefont {J.-Z.}\ \bibnamefont {{Qi}}}, \bibinfo {author} {\bibfnamefont {J.-W.}\ \bibnamefont {{Zhao}}}, \bibinfo {author} {\bibfnamefont {S.}~\bibnamefont {{Cao}}}, \bibinfo {author} {\bibfnamefont {M.}~\bibnamefont {{Biesiada}}}, \ and\ \bibinfo {author} {\bibfnamefont {Y.}~\bibnamefont {{Liu}}},\ }\href {\doibase 10.1093/mnras/stab638} {\bibfield  {journal} {\bibinfo  {journal} {\mnras}\ }\textbf {\bibinfo {volume} {503}},\ \bibinfo {pages} {2179} (\bibinfo {year} {2021})},\ \Eprint {http://arxiv.org/abs/2011.00713} {arXiv:2011.00713 [astro-ph.CO]} \BibitemShut {NoStop}%
\bibitem [{\citenamefont {Wong}\ \emph {et~al.}(2020)\citenamefont {Wong}, \citenamefont {Suyu}, \citenamefont {Chen}, \citenamefont {Rusu}, \citenamefont {Millon}, \citenamefont {Sluse}, \citenamefont {Bonvin}, \citenamefont {Fassnacht}, \citenamefont {Taubenberger}, \citenamefont {Auger} \emph {et~al.}}]{Wong:2019kwg}%
  \BibitemOpen
  \bibfield  {author} {\bibinfo {author} {\bibfnamefont {K.~C.}\ \bibnamefont {Wong}}, \bibinfo {author} {\bibfnamefont {S.~H.}\ \bibnamefont {Suyu}}, \bibinfo {author} {\bibfnamefont {G.~C.}\ \bibnamefont {Chen}}, \bibinfo {author} {\bibfnamefont {C.~E.}\ \bibnamefont {Rusu}}, \bibinfo {author} {\bibfnamefont {M.}~\bibnamefont {Millon}}, \bibinfo {author} {\bibfnamefont {D.}~\bibnamefont {Sluse}}, \bibinfo {author} {\bibfnamefont {V.}~\bibnamefont {Bonvin}}, \bibinfo {author} {\bibfnamefont {C.~D.}\ \bibnamefont {Fassnacht}}, \bibinfo {author} {\bibfnamefont {S.}~\bibnamefont {Taubenberger}}, \bibinfo {author} {\bibfnamefont {M.~W.}\ \bibnamefont {Auger}},  \emph {et~al.},\ }\href@noop {} {\bibfield  {journal} {\bibinfo  {journal} {Monthly Notices of the Royal Astronomical Society}\ }\textbf {\bibinfo {volume} {498}},\ \bibinfo {pages} {1420} (\bibinfo {year} {2020})}\BibitemShut {NoStop}%
\bibitem [{\citenamefont {Shajib}\ \emph {et~al.}(2020)\citenamefont {Shajib} \emph {et~al.}}]{shajib2019strides}%
  \BibitemOpen
  \bibfield  {author} {\bibinfo {author} {\bibfnamefont {A.}~\bibnamefont {Shajib}} \emph {et~al.} (\bibinfo {collaboration} {DES}),\ }\href {\doibase 10.1093/mnras/staa828} {\bibfield  {journal} {\bibinfo  {journal} {Mon. Not. Roy. Astron. Soc.}\ }\textbf {\bibinfo {volume} {494}},\ \bibinfo {pages} {6072} (\bibinfo {year} {2020})},\ \Eprint {http://arxiv.org/abs/1910.06306} {arXiv:1910.06306 [astro-ph.CO]} \BibitemShut {NoStop}%
\bibitem [{\citenamefont {{Luongo}}\ and\ \citenamefont {{Muccino}}(2020)}]{2020A&A...641A.174L}%
  \BibitemOpen
  \bibfield  {author} {\bibinfo {author} {\bibfnamefont {O.}~\bibnamefont {{Luongo}}}\ and\ \bibinfo {author} {\bibfnamefont {M.}~\bibnamefont {{Muccino}}},\ }\href {\doibase 10.1051/0004-6361/202038264} {\bibfield  {journal} {\bibinfo  {journal} {\aap}\ }\textbf {\bibinfo {volume} {641}},\ \bibinfo {eid} {A174} (\bibinfo {year} {2020})},\ \Eprint {http://arxiv.org/abs/2010.05218} {arXiv:2010.05218 [astro-ph.CO]} \BibitemShut {NoStop}%
\end{thebibliography}

%

\end{document}